\documentclass[12pt]{article}
\pdfoutput=1
\usepackage{amssymb, amsmath}
\usepackage{cite}
\usepackage{graphicx}
\usepackage{xcolor}
\usepackage{hyperref}
\definecolor{nicered}{rgb}{0.7,0.1,0.1}
\definecolor{nicegreen}{rgb}{0.1,0.5,0.1}
\hypersetup{colorlinks,citecolor= nicegreen,linkcolor= nicered}
\newcommand{\be}  {\begin{equation}}
\newcommand{\ee}  {\end{equation}}

\begin{document}
\begin{titlepage}
  \newcommand{\AddrIPM}{{\sl \small School of physics, Institute for Research in Fundamental Sciences (IPM)\\
  \sl \small P.O. Box 19395-5531, Tehran, Iran}}
  \newcommand{\MP}{{\sl \small Max-Planck-Institut f\"ur Kernphysik\\
  \sl \small Saupfercheckweg 1, 69117 Heidelberg, Germany\\ \sl \small and \\
  \sl \small International Centre for Theoretical Physics, ICTP\\
  \sl \small Strada Costiera 11, 34151 Trieste, Italy}}
  \newcommand{\AddrNYU}{{\sl \small New York University Abu Dhabi\\
  \sl \small P.O. Box 129188, Saadiyat Island, Abu Dhabi, United Arab Emirates}}
  \vspace*{0.5cm}
\begin{center}
  \textbf{\boldmath \large Neutrinos from GRB 221009A:\\producing ALPs and explaining LHAASO anomalous $\gamma$ event}
  \vspace{0.8cm}\\
   Nicol\'as Bernal\footnote{\href{mailto:nicolas.bernal@nyu.edu}{\tt nicolas.bernal@nyu.edu}}\\
\AddrNYU.
\vspace{0.4cm}\\
 Yasaman Farzan\footnote{\href{mailto:yasaman@theory.ipm.ac.ir}{\tt yasaman@theory.ipm.ac.ir}} \\
  \AddrIPM.   \vspace*{0.4cm}\\
   Alexei Yu. Smirnov\footnote{\href{mailto:smirnov@mpi-hd.mpg.de}{\tt smirnov@mpi-hd.mpg.de}}\\
  \MP.
  \vspace{1cm}\\
\end{center}

\begin{abstract}
We propose a novel explanation for the 18~TeV gamma ray from GRB 221009A  observed by LHAASO. High-energy neutrinos are converted into axion-like particles (ALPs) via their interaction with the cosmic neutrino background. Subsequently, ALPs are converted into high-energy photons in the magnetic field of our galaxy. We compute the fluxes of neutrinos, ALPs, and photons reaching Earth. IceCube's constraints on the neutrino flux from GRB 221009A translate into a severe upper bound on the photon flux. We find a range of parameters where all existing bounds are satisfied and the 18~TeV LHAASO photon can be explained. In the future, the specific correlation between the photon and neutrino flux reaching Earth from powerful neutrino sources with energies larger than 10~TeV such as GRBs or AGNs, can be used as a tool to differentiate our explanation from the alternatives suggested in the literature. We discuss how the interactions of our scenario can be embedded within electroweak gauge-invariant models, avoiding various cosmological and terrestrial bounds. We comment on the possibility of explaining the 251~TeV photon observed by the Carpet-2 detector, taking into account the bounds from the observation of high-energy neutrinos from TXS 0506+056. 
\end{abstract}

\end{titlepage}
\setcounter{footnote}{0}

\section{Introduction} \label{sec:intro}
Gamma-Ray Bursts (GRBs) were first discovered in 1967 and since then have puzzled the scientific community by their high luminosity~\cite{Klebesadel:1973iq}. So far, more than 1700 GRB events have been found, the most luminous among them being GRB 221009A detected first by Fermi-GBM~\cite{2022GCN.32636....1V, Lesage:2023vvj} on the 9$^\text{th}$ of October 2022 and then by other detectors.
The very high-energy gamma rays from GRB 221009A observed by the Large High-Altitude Air Shower Observatory (LHAASO)~\cite{2022GCN.32677....1H} and by Carpet-2~\cite{2022ATel15669....1D} defy an explanation within the Standard Model of particle physics (SM).
A $\gamma$ flux at 18~TeV is expected to be strongly attenuated by $e^-e^+$ pair production through scattering off background photons while traveling from the source at a redshift of $z = 0.15$ to Earth.  
In fact, for $\gamma$ with $E_\gamma = 18$~TeV, the optical depth is estimated to be $\tau_\gamma \simeq 15$~\cite{Baktash:2022gnf}. For LHAASO, the precision of the $\gamma$ energy reconstruction is $36\%$~\cite{Ma:2022aau}, that is, $E_\gamma = 18 \pm 6.5$~TeV, but even for $E_\gamma = 10$~TeV, the optical depth is $\tau_\gamma \simeq 5$ so an enormous initial $\gamma$-ray flux is required to explain this event. Indeed, it is challenging to produce such high-energy photons inside GRBs via synchrotron-self Compton scenario which is the most common model for the emission of GRBs~\cite{Rojas:2023jdd}. Detection of this high-energy photon may be a herald of new physics.

Several beyond-SM explanations have been suggested since the observations of these events. One group of explanations is based on the conversion of high-energy photons from the source to axions and the reconversion of axions to photons in the magnetic field of the Milky Way~\cite{Galanti:2022pbg, Baktash:2022gnf, Lin:2022ocj, Troitsky:2022xso, Nakagawa:2022wwm, Gonzalez:2022opy, Rojas:2023jdd}. However, Ref.~\cite{Carenza:2022kjt} demonstrates that if one extrapolates the low-energy ($0.1~{\rm GeV} \lesssim E_\nu \lesssim 1$~GeV) photon flux measured by Fermi-LAT to higher energies at the source, the explanation of the 18~TeV LHAASO event requires a photon-axion coupling $g_{a \gamma \gamma}$ already in tension with magnetic white dwarf polarization~\cite{Marsh:2017yvc, Fortin:2018ehg, Reynolds:2019uqt, Fortin:2021sst, Reynes:2021bpe, Dessert:2022yqq}.

Another set of explanations is based on the production at the source of heavy high-energy neutrinos (with masses in the MeV range), which then undergo a radiative decay on their way to Earth~\cite{Cheung:2022luv, Smirnov:2022suv, Brdar:2022rhc}.
In Ref.~\cite{Cheung:2022luv} these neutrinos are produced via the conversion of SM neutrinos to the heavy ones with large off-diagonal magnetic moments in the magnetic field of GRB 221009A~\cite{Huang:2022udc}. 
Alternatively, in Refs.~\cite{Smirnov:2022suv} and~\cite{Brdar:2022rhc}, the MeV-scale neutrinos are produced in decays of pions and kaons by mixing with SM neutrinos. This scenario is strongly constrained by the BBN bound and by the gamma flux of SN 1987A~\cite{Huang:2022udc, Guo:2023bpo}.
Another explored possibility includes effects of Lorentz violation which increase with energy~\cite{Baktash:2022gnf, Li:2022wxc, Finke:2022swf, Li:2023rgc}.

In this paper, we propose a novel explanation that employs the interaction of SM neutrinos with ALPs. In contrast to the axion solutions proposed earlier, here ALPs are produced via the interactions of high-energy neutrinos from GRBs with neutrinos of the cosmic neutrino background (C$\nu$B) rather than by conversion of $\gamma$ in the magnetic field of the host galaxy. The ALP flux is generated all the way from the source to Earth. Then, as in previous axionic explanations, ALPs are converted into photons in the magnetic field of our galaxy. The key feature of our scenario is that the neutrino flux from remote sources is absorbed in annihilation with relic neutrinos. Searches for neutrinos from GRB 221009A itself and observation of neutrinos from the TXS 0506+056 blazar~\cite{IceCube:2018cha} put strong bounds on the scenario.  Unlike the axionic explanations previously proposed in the literature, our explanation works even if the photon ALP coupling, $g_{a \gamma \gamma}$ is smaller than the current bounds by a factor of 4. Moreover, in this model, we do not need the synchrotron-self Compton mechanism to produce high-energy photons. The initial high-energy neutrinos are produced by canonical hadronic processes. We discuss how the scenario can be accommodated within a viable gauge-invariant model satisfying all terrestrial, astrophysical, and cosmological bounds. Our scenario predicts the detection of high-energy photon flux accompanying the high-energy neutrino flux from other sources at a cosmological distance, such as active galactic nuclei (AGNs).

The paper is organized as follows. In Section~\ref{sec:NuALP}, we describe our scenario and compute the rates of processes relevant to the evolution of the ALP and neutrino fluxes. 
In Section~\ref{sec:ALPsNUs}, we compute the neutrino and ALP fluxes as functions of redshift from the source to Earth, using the evolution equations detailed in Appendix~\ref{app:Evolution}. In section~\ref{sec:bounds}, we first consider the implications of the IceCube bound on the neutrino flux from GRB 221009A as well as of the observation of neutrinos from the point source TXS 0506+056. We then discuss the astrophysical and cosmological bounds. In Section~\ref{sec:GRB}, we first show how our scenario can explain the LHAASO 18~TeV event and discuss methods to test this explanation and distinguish it from alternative explanations.   We then examine the possibility of explaining the Carpet-2 251~TeV event. In Appendix~\ref{app:UVcompletion}, we give examples of UV-complete models that accommodate our scenarios and discuss how various existing bounds can be avoided. In Appendix~\ref{app:ALPconversion}, we present formulas for the ALP photon conversion and discuss why an excess of photons in the GeV range due to the ALPs in our model is not expected.
Our results are summarized in Section~\ref{sec:Conclusions}.

\section{Neutrino - ALP scenarios\label{sec:NuALP}}
GRBs are powerful sources of cosmic rays as well as neutrinos with energies up to the multi-PeV scale.
During their propagation in the interstellar medium, these neutrinos can annihilate with neutrinos of the C$\nu$B, producing a flux of ALPs. 
Subsequently, a fraction of ALPs converts to photons in the magnetic field of our galaxy and can be observed as a flux of $\gamma$ rays by LHAASO.  
In our proposal, ALPs, denoted by $a$, are meta-stable particles with a mass $m_a \lesssim {\rm few} \times 10^{-7}$~eV that interact with photons through the coupling
\be \label{aFF}
    \frac{g_{a\gamma\gamma}}{4}\, a\, F^{\mu\nu} \tilde{F}_{\mu\nu}\,.
\ee
Here $F$ is the electromagnetic field strength tensor, $\tilde F$ is its dual, and $g_{a\gamma\gamma}$ is a coupling with a dimension of inverse energy.

Two scenarios for neutrino $\nu$ annihilation into ALPs are considered: $i)$ $\nu\, \nu\to a\, a$ and $ii)$ $\nu\, \nu \to a\, a'$ in which $a'$ is another
scalar particle with a mass of 50~eV $\lesssim m_{a'}\ll 50$~keV, that later decays back to $a\, \nu\, \nu$ or $a\, \bar{\nu}\, \bar{\nu}$. The need for these mass ranges will be explained later. Let us first briefly discuss the important features of each scenario.
\begin{itemize} 
    \item {\it The $\nu\, \nu \to a\, a$ scenario:}
    Neutrino annihilation into a pair of ALPs proceeds through the effective interaction 
    \be 
        \label{aanunu} 
        \frac{1}{4\, \Lambda_{aa}}\, a^2\, \nu^T\, c\, \nu\,,
    \ee
    where $c$ is an asymmetric $2\times 2$ matrix with off-diagonal elements of $\pm 1$ acting on the spinorial indices.
    A possible UV completion of this effective term is presented in Appendix~\ref{app:UVcompletion}.
    
    In the early Universe, the interaction in Eq.~\eqref{aanunu} can bring ALPs to thermal equilibrium with the SM bath, violating strong bounds on the number of effective relativistic degrees of freedom $\Delta N_\text{eff}$. In order to prevent the relativistic ALP production in the early Universe and therefore to respect the BBN and CMB bounds on $\Delta N_\text{eff}$, we shall introduce a mechanism based on a varying mass for the mediator induced by its coupling to dark matter field in Appendix~\ref{app:UVcompletion}. Even if this mechanism prevents the production of $a$ in the early Universe, at late times, $\nu\, \nu \to a\, a$ and $\bar\nu\, \bar\nu \to a\, a$ processes populate the ALP background.

    A flux of high-energy ALPs is produced not only by the scattering of high-energy neutrinos off of the C$\nu$B ($\nu\, \nu \to a\, a$ and $\bar\nu\, \bar\nu \to a\, a$), but also by elastic scatterings of neutrinos off the ALP background: $\nu\, a \to \nu\, a$ and $\bar\nu\, a \to \bar\nu\, a$.
    In turn, high-energy ALPs can lose energy by scattering off the C$\nu$B ($a\, \nu \to a\, \bar{\nu}$ or $a\, \bar{\nu} \to a\, \nu$) or even annihilating with background ALPs ($a\, a\to \nu\, \nu$ and $\bar{\nu}\, \bar{\nu}$). The relevant cross sections are computed below. The evolution of neutrino and ALP fluxes is presented in Section~\ref{sec:ALPsNUs} and in Appendix~\ref{app:Evolution}.

    \item {\it The $\nu\, \nu\to a\, a'$ scenario:} 
    The effective interaction leading to annihilation $\nu\, \nu\to a\, a'$ is 
    \be 
        \label{aaprimenunu} 
        \frac{1}{2\, \Lambda_{aa'}}\, a\, a'\, \nu^T\, c\, \nu\,,
    \ee
    with a possible UV completion also discussed in Appendix~\ref{app:UVcompletion}. When the temperature of the Universe drops below the $a'$ mass, the background neutrinos cannot produce $a\, a'$ pairs. Thus, if we introduce a mechanism that prevents ALP production in the early Universe when the temperature is above the $a'$ mass, the density of the ALP background today can be low. An example of such a mechanism is presented in Appendix~\ref{app:UVcompletion}. As a result, the scattering of the high-energy neutrinos or ALPs off the ALP relic background becomes negligible. While the processes that lead to the production of high-energy ALP flux are only $\nu\, \nu$, $\bar{\nu}\, \bar{\nu} \to a\, a'$,  the processes that suppress the flux are $a\, \nu\to a'\, \bar{\nu}$ and $a\, \bar{\nu} \to a'\, \nu$.
\end{itemize}

Let us now discuss the flavor structure of the couplings. Unlike $\nu_\tau$, the new interactions of $\nu_e$ and $\nu_\mu$ are substantially constrained by various laboratory experiments. Throughout this paper, we set the couplings $\nu_e$ and $\nu_\mu$ to zero and only turn on the coupling to $\nu_\tau$. Such $\tau$-philic and $e$- and $\mu$-phobic models can be justified by a simple $U(1)$ flavor symmetry (see Appendix~\ref{app:UVcompletion} for a concrete example).
As is well known, both the relic neutrinos today and the high-energy neutrinos from sources at cosmic distances are in the form of incoherent mass eigenstates. 
In terms of mass eigenstates, the interactions in Eqs.~\eqref{aanunu} and~\eqref{aaprimenunu} can be rewritten as
\begin{align}
    \frac{1}{4\, \Lambda_{aa}}\, a^2\, \nu_\tau^T\, c\, \nu_\tau & = \sum_{i,j} U_{\tau i}\, U_{\tau j}\, \frac{(\nu_i^T\, c\, \nu_j)\, a^2}{4\, \Lambda_{aa}}\,, \label{eq:aa1}\\
    \frac{1}{2\, \Lambda_{aa'}}\, a\, a'\, \nu_\tau^T\, c\, \nu_\tau & = \sum_{i,j} U_{\tau i}\, U_{\tau j}\, \frac{(\nu_i^T\, c\, \nu_j)\, a\, a'}{2\, \Lambda_{aa'}}\, . \label{eq:aaprime}
\end{align}

It is important to note that in the mass basis, there are off-diagonal interaction terms that can in principle lead to decays of the heavier neutrinos to the lighter ones. For example, with the interaction term in Eq.~\eqref{eq:aa1} and $\Lambda_{aa}\sim 1$~GeV, the lifetimes of the heavier neutrinos at rest become much shorter than the age of the Universe. Thus, the relic $\nu_3$ and $\nu_2$ have already decayed into $\nu_1$. For definiteness, here we assume the normal mass hierarchy, which implies that $m_3 \simeq \sqrt{\Delta m_\text{atm}^2}$ and $m_2 \simeq \sqrt{\Delta m_\text{sol}^2}$, where $\Delta m_\text{atm}^2$ and $\Delta m_\text{sol}^2$ are the atmospheric and solar mass-squared differences, respectively, and $m_i$ corresponds to the neutrino mass eigenvalues. Then in the $\nu\, \nu \to  a\, a$ scenario, the interaction in Eq.~\eqref{eq:aa1} leads to the decay rates at rest 
\be \label{tau3}
    \Gamma (\nu_3\to \bar{\nu}_{2,1}\, a\, a)\sim \frac{m_3^3}{64\pi^3\, \Lambda_{aa}^2}\, |U_{\tau 3}|^2\, \left(1-|U_{\tau 3}|^2\right) \simeq 10^{-3} \left(\frac{1~{\rm GeV}}{\Lambda_{aa}}\right)^2 {\rm year}^{-1},
\ee 
and 
\be \label{tau2}
    \Gamma (\nu_2\to \bar{\nu}_1\, a\, a) \sim \frac{m_2^3}{64\pi^3\, \Lambda_{aa}^2}\, |U_{\tau 2}|^2\, |U_{\tau 1}|^2 \simeq 10^{-6} \left(\frac{1~{\rm GeV}}{\Lambda_{aa}}\right)^2 {\rm year}^{-1} . 
\ee 
The strongest bound on invisible neutrino decay comes from CMB~\cite{Barenboim:2020vrr, Chen:2022idm} (see also Refs.~\cite{Hannestad:2005ex, Basboll:2008fx, Escudero:2019gfk}). Lifetimes in Eqs.~\eqref{tau3} and~\eqref{tau2} satisfy even the most stringent bound; however, the lifetimes of $\nu_2$ and $\nu_3$ are predicted to be much shorter than the age of the Universe. Thus, the relic $\nu_2$ and $\nu_3$ should now have decayed.
Therefore, we shall end up only with the  $\nu_1$ and $\bar{\nu}_1$  backgrounds. High-energy ALPs with a boost factor $\sim 10^{14}\, (E_a/10~{\rm MeV})\, (10^{-7}~{\rm eV}/m_a)$ have a lifetime long enough to survive from far galaxies at high redshift to Earth.

In the $\nu\, \nu \to a\, a'$ scenario with $m_{a'} \gg m_3$ three-body decays of SM neutrinos are kinematically forbidden, but 5-body decay modes $\nu_3 \to \bar{\nu}_{1,2}\, a^2\, \nu_{1,2}\, \nu_{1,2}$ and $\nu_2 \to \bar{\nu}_{1}\, a^2\, \nu_{1}\, \nu_{1}$ through loops are allowed. However, their decay rates are suppressed by factors $(m_i/m_{a'})^2\, (m_j/\Lambda_{aa'})^2 / (16\pi^2)^2$, so that the lifetimes are larger than the age of the Universe. Overall, while in the $\nu\, \nu \to a\, a$ scenario only the $\nu_1$ and $\bar{\nu}_1$ relic background remains today, in the $\nu\, \nu \to a\, a'$ scenario the C$\nu$B consists of all three neutrino mass eigenstates in equal amounts, just as in the SM. 

Now, let us focus on the interaction rates. The center-of-mass energy $\sqrt{s}$ of a neutrino or ALP of energy $E$ and a non-relativistic $\nu_3$ is
\begin{equation}
    \sqrt{s} \simeq \sqrt{2\, E\, m_3} \simeq \sqrt{2\, E\, \sqrt{\Delta m_\text{atm}^2}} \simeq 2.2~{\rm MeV}\, \sqrt{\frac{E}{40~{\rm TeV}}}\,.
\end{equation}
For scatterings with relativistic $\nu_1$ of energy $\sim 10^{-4}$~eV, one has $\sqrt{s} \sim 50$~keV. Therefore, as long as the mediator of the effective interaction in Eq.~\eqref{aanunu} is heavier than $\sim 10$~MeV, we can safely use the effective interaction formalism. As long as $m_a, m_{a'} \ll 50$~keV, the center-of-mass energy is large enough to produce a pair $a\, a'$. Using the interaction~\eqref{eq:aa1}, we find that  independently of the energy of the initial particles
\be \label{sigma}
    \sigma_{ij}\equiv \sigma(\nu_i\nu_j \to aa) =  \sigma(\bar{\nu}_i\bar{\nu}_j \to aa)= \sigma(aa\to \nu_i\nu_j )=  \sigma(aa\to \bar{\nu}_i\bar{\nu}_j) \simeq \frac{|U_{\tau i}|^2\, |U_{\tau j}|^2}{32 \pi\, \Lambda_{aa}^2}\,,
\ee
with $U_{\alpha i}$ being the elements of the PMNS matrix.
Since the energy $E_\text{fin}$ of the final particle is uniformly distributed between zero and the energy of the incident particle $E_\text{ini}$, we obtain the differential cross section as
\be \label{diffnunu}
    \frac{d\sigma}{dE_\text{fin}} = 2\, \sigma\, \frac{1}{E_\text{ini}}\,,
\ee
for $0<E_\text{fin}<E_\text{ini}$. The factor of 2 reflects the symmetry of identical final states. Similarly, 
\be \label{sigmas}
    (\sigma_{s} )_{ij}\equiv \sigma(a  \nu_i \to a  \bar{\nu}_j) = \sigma(a  \bar{\nu}_i \to a  {\nu}_j) \simeq \frac{|U_{\tau i}|^2\, |U_{\tau j}|^2}{32\pi\, \Lambda_{aa}^2},
\ee
where the subscript $s$ stands for ``scattering''. The energy distribution of the final $a$ is given by
\be \label{diffsig}
    \frac{d \sigma_s}{dE_\text{fin}} = 2\sigma_s \frac{1}{E_\text{ini}} \left(1-\frac{E_\text{fin}}{E_\text{ini}}\right), 
\ee
again for $0<E_\text{fin}<E_\text{ini}$.
On the other hand, neglecting the $a'$ mass, from Eq.~\eqref{eq:aaprime}, one obtains  
\be \label{sigma'}
    \sigma_{ij}^{\prime}\equiv \sigma(\nu_i\nu_j \to aa')=  \sigma(\bar{\nu}_i\bar{\nu}_j \to aa') \simeq  \frac{|U_{\tau i}|^2\, |U_{\tau j}|^2}{16 \pi\, \Lambda_{aa'}^2}\,,
\ee
and
\begin{equation}
    \frac{d\sigma^\prime}{dE_\text{fin}}= \sigma^\prime \frac{1}{E_\text{ini}}\,.
\end{equation}
Note that the difference between Eqs.~\eqref{sigma} and~\eqref{sigma'} is a factor of 2 which again reflects the symmetry factor of the final states. The scattering cross section $(\sigma')_{ij} =\sigma (a\, \nu_i \to \bar{\nu}_j\, a')=\sigma (a\, \bar{\nu}_i \to {\nu}_j\, a')$ is given by Eq.~\eqref{sigmas}, replacing $\Lambda_{aa} \to \Lambda_{aa'}$. 

In standard cosmology, we expect the densities of all relic neutrino and antineutrino mass eigenstates to be equal
\be \label{nStandard}
    n_1=n_2=n_3=n_{\bar{1}}=n_{\bar{2}}=n_{\bar{3}}=56 ~{\rm cm}^{-3}(1+z)^3 \ .
\ee
If ALPs were in thermal equilibrium with neutrinos before the neutrino decoupling era, their number density $n_a$ currently equals $n_a = 4\, n_i/3$. In the $\nu\, \nu \to a\, a$ scenario, if there is a mechanism that prevents ALP production before recombination, the entropy of the neutrino bath in the late times will be shared with the ALPs due to decays of $\nu_2$ and $\nu_3$. Thus, we would have 
\begin{align} \label{nnn}
    &n_3=n_2= n_{\bar{3}}=n_{\bar{2}}=0 ,\\
    &n_1= n_{\bar{1}} \simeq 56 ~{\rm cm}^{-3}(1+z)^3\, \frac{3\times 2 \times 7/8}{ 2 \times 7/8+1} = 107~{\rm cm}^{-3}\, (1+z)^3, 
\end{align}
and again $n_a = 4\, n_1/3$. 
In the $\nu\, \nu \to a\, a'$ scenario, $\nu_3$ and $\nu_2$ are stable and ALPs cannot be produced late by the annihilation and decay of relic neutrinos. Thus, if we implement a mechanism to prevent ALP production in the early Universe when the temperature is above $m_{a'}$, the background ALP density remains zero $n_a=0$, and the relic neutrinos have standard densities as in Eq.~\eqref{nStandard}.

For the $\nu\nu\to aa$ scenario, we can define the total absorption rate for $\nu_i$ and $\bar{\nu}_i$ as
\be \label{Gammi}
    \Gamma_i = \sum_j \sigma_{ij}\, n_j + \sum_j (\sigma_s)_{ij}\, n_a \simeq \frac{|U_{\tau i}|^2}{32\pi\, \Lambda_{aa}^2}\, n_1 \left(|U_{\tau 1}|^2 + \frac43\right),
\ee
while for the $\nu\, \nu \to a\, a'$ scenario,
\be \label{Gammiprime}
    \Gamma_i^\prime = \sum_j \sigma_{ij}^\prime\, n_j \simeq \frac{|U_{\tau i}|^2}{16\pi\, \Lambda_{aa'}^2}\, n_1\,.
\ee 
Similarly, we can define a rate for the scattering of ALPs off the background. With the effective interaction~\eqref{eq:aa1}, the ALP scattering rate is
\be \label{GammaALP}  
    \Gamma_a = 2\, n_a \sum_{i, j} \sigma_{ij} + \sum_{i,j}(n_i + n_{\bar{i}})\, (\sigma_s)_{ij} \simeq \frac{n_1\, |U_{\tau 1}|^2 + n_a}{16 \pi\, \Lambda_{aa}^2} = \frac{2}{|U_{\tau i}|^2}\, \Gamma_i\,,
\ee
where the factor of 2 in front of $n_a$ reflects the fact that the ALP pair can annihilate both into a neutrino pair and an antineutrino pair.
For the effective interaction~\eqref{eq:aaprime}, we find a similar total rate for the $a$ scattering off the background
\be \label{GammaALPrime}
    \Gamma_a'=\sum_{i,j}(n_i+n_{\bar{i}})\, (\sigma_s')_{ij} = \frac{2}{|U_{\tau i}|^2}\, \Gamma'_i\,.
\ee
In Appendix~\ref{app:Evolution}, we derive the evolution equations for high-energy neutrino and ALP fluxes on cosmological distances.

Finally, because of the interaction~\eqref{aFF}, an ALP in the magnetic field can be converted into a photon. For $m_a \lesssim 3 \times 10^{-7}$~eV and for energies greater than a few TeVs, the conversion probability is approximately constant in energy and can be parameterized as~\cite{Carenza:2022kjt}
\be \label{Pagamma}
    P_{a\to \gamma}\simeq 1.5\times 10^{-4} \left( \frac{g_{a\gamma \gamma}}{3\times 10^{-12}~{\rm GeV}^{-1}}\right)^2.
\ee
 As shown in Appendix~\ref{app:ALPconversion}, for low energies $E \sim [0.1,\, 1]$~GeV relevant for Fermi-LAT,  the probability $ P_{a\to \gamma}$ is severely suppressed. Using the formulas in Appendix~\ref{app:ALPconversion}, we observe that for $m_a$ larger than a few~$\times 10^{-7}$~eV, the conversion probability is suppressed. For $m_a \lesssim 10^{-11}$~eV, there are more stringent limits on $g_{a \gamma \gamma}$~\cite{Marsh:2017yvc, Reynolds:2019uqt, Reynes:2021bpe, Dessert:2022yqq}. Therefore, we take $m_a$ in  the range $\sim [10^{-11},\, 10^{-7}]$~eV.
  
\section{ALP and neutrino fluxes} \label{sec:ALPsNUs}
The evolution of the neutrino Flux  $F_{\nu_i}$ with the redshift  is given by equation
\begin{align} \label{Fnui0}
	&\frac{d \tilde{F}_{\nu_i}(z,E_\nu)}{d z}= \frac{ {\tilde{F}_{\nu_i}}(z,E_\nu)}{1+ z}	- \frac{\partial \tilde{F}_{\nu_i}(z,E_\nu)}{\partial E_\nu}\frac{E_\nu}{(1+z)^2} \nonumber\\
	&\quad+ \left(- \Gamma_i \tilde{F}_{\nu_i}(z,E_\nu)+2\int_{E_{\nu}}\sum_j n_j(z) v \frac{d\sigma_s(a(E_a)+\bar{\nu}_j \to a+\nu_i(E_{\nu}))}{d E_{\nu}}\tilde{F}_a(z,E_a)dE_a\right) \frac{dt}{dz}\, \nonumber \\
	&\quad+ \left(2\int_{E_{\nu}}\sum_j n_a v \frac{d\sigma(a(E_a)+a \to {\nu}_j+\nu_i(E_{\nu}))}{d E_{\nu}}\tilde{F}_a(z,E_a)dE_a\right) \frac{dt}{dz}\nonumber\\&
	\quad+ \left(\int_{E_{\nu}}\sum_j n_a v \frac{d\sigma_s(\bar{\nu}_j(E_{\nu_j})+a \to a+\nu_i(E_\nu))}{d E_{\nu}}\tilde{F}_{\nu_j}(z,E_{\nu_j})dE_{\nu_j}\right) \frac{dt}{dz} \,,
\end{align}
where the factors of 2 in the second and third lines are there because we consider the sum of neutrino and antineutrino fluxes. Similarly for the ALPs flux,
\begin{align} \label{Fa0}
	&\frac{d \tilde{F}_a(z,E_a)}{d z}= \frac{\tilde{F}_a(z,E_a)}{1+ z}	- \frac{\partial \tilde{F}_a(z,E_a)}{\partial E_a}\frac{E_a}{(1+z)^2} \nonumber\\
	&\qquad+ \frac{dt}{dz} \left(- \Gamma_a \tilde{F}_a(z,E_a)+  \int_{E_a}\sum_{i,j} n_j(z) v \frac{d\sigma(\nu_i(E_{\nu_i})+\nu_j \to a(E_a)+a)}{d E_a}\tilde{F}_{\nu_i}(z,E_{\nu_i})dE_{\nu_i}\right) \nonumber\\
	&\qquad+ 2\frac{dt}{dz} \left( \int_{E_a}\sum_j n_j(z) v \frac{d\sigma(a(E_a')+\nu_j \to a(E_a)+\bar{\nu}_i)}{d E_a}\tilde{F}_a(z,E_a')dE_a' \right)\nonumber \\
	&\qquad+ \frac{dt}{dz} \left( \int_{E_a}\sum_{i,j} n_a(z) v \frac{d\sigma(a+\nu_i(E_{\nu_i })\to a(E_a)+\bar{\nu}_j)}{d E_a}\tilde{F}_{\nu_i}(z,E_{\nu_i})dE_{\nu_i} \right),
\end{align}
where $\tilde{F}_a \equiv F_a\, r^2(z=0)$ and $\tilde{F}_{\nu_i} \equiv F_{\nu_i}\, r^2(z=0)$, with $r$ being the physical distance to the source and $E_\nu$ ($E_a$) is the energy of the neutrino (ALP).
All details of the derivation of the evolution equations for the ALP and neutrino fluxes are given in Appendix~\ref{app:Evolution}.
Equations~\eqref{Fnui0} and~\eqref{Fa0} have to be solved with the initial conditions
\be \label{initial}
    \tilde{F}_a(z=0.15, E_a) = 0 \qquad {\rm and} \qquad \tilde{F}_{\nu_i}(z=0.15, E_{\nu_i}) = \mathcal{\tilde{A}}\, E_{\nu_i}^{-\kappa}\,.
\ee

The terms in the first lines of Eqs.~\eqref{Fnui0} and~\eqref{Fa0} are related to the transition from the equation for coordinates to the redshift and take into account the expansion of the Universe. The first term of the second line of Eq.~\eqref{Fnui0} corresponds to the scattering of the $\nu$ flux off the relic background as given by Eq.~\eqref{Gammi}. The following terms in Eq.~\eqref{Fnui0}, which we collectively call neutrino regeneration terms, describe the production of high-energy neutrinos due to scatterings of the high-energy ALP and neutrino fluxes off the background neutrinos and ALPs. In Eq.~\eqref{Fa0}, the first term on the second line gives a reduction in flux due to scattering off the background particles with $\Gamma_a$ given by Eq.~\eqref{GammaALP}. The next term accounts for the production of ALPs due to the scattering of high-energy neutrinos off the C$\nu$B. Regeneration of ALPs after scattering off background neutrinos is given by the term in the third line of Eq.~\eqref{Fa0} which is called the ALP regeneration term. The last term accounts for boosting a background ALP to high energies via the interactions of high-energy neutrino flux.

For the evolution equations of the fluxes in the $\nu\, \nu \to a\, a'$ scenario, we can rewrite Eqs.~\eqref{Fnui0} and~\eqref{Fa0}  replacing $\Gamma_i$, $\Gamma_a$ and the differential cross sections with the corresponding primed quantities given in Eqs.~\eqref{sigma'}, \eqref{Gammiprime}, and~\eqref{GammaALPrime}, setting $n_a=0$ and using densities as shown in Eq.~\eqref{nStandard}.

Note that in the present analysis the scattering angles have not been considered. The typical scattering angle is $\sqrt{s}/E_\gamma \sim \mathcal{O}(1)~\text{MeV} / (\mathcal{O}(10)~\text{TeV}) \sim 10^{-7}$, which is much smaller than the opening angle of the jet inside the GRB, and therefore validates our simplification.

The solution to Eqs.~\eqref{Fnui0} and~\eqref{Fa0} can be in general parameterized as 
\be \label{FNUparamet}
    \tilde{F}_{\nu_i}(z, E_\nu) = \tilde{\mathcal{A}}\, E_\nu^{-\kappa}\, \exp\left[g_i(z , E_\nu)\right], 
\ee
where $\kappa = 1.7 - 2.5$ with reference value $\kappa = 2$.
The $g_i$ functions take care of the redshift effect, as well as the scattering of the neutrino {\it en route} to Earth. From Eq.~\eqref{initial}, $g(z=0.15, E_\nu) = 1$.
In the following, we show that the neutrino and ALP regeneration terms (that is, the last three terms in Eq.~(\ref{Fnui0}) and the term in the third line of Eq.~\eqref{Fa0}) have only a subdominant effect. If we neglect the regeneration of the high-energy neutrinos, the equation for neutrino flux decouples. Its solution can then be written as shown in Eq.~\eqref{FNUparamet} with
\be \label{g}
    g_i(z, E_\nu) \equiv \ln\left[\frac{1+z}{1 + z_L}\right] - \kappa \left(\frac{1}{1+z} - \frac{1}{1+ z_L}\right) - \int_z^{z_L}\frac{\Gamma_i^{(\prime)}}{H\, (1+z')}\,dz'\, ,
\ee
where $z_L$ is the redshift of the source which, for GRB 221009A, is $z_L = 0.15$.  The last term is the optical depth term in which $\Gamma_i$  ($\Gamma_i'$) accounts for the neutrino scattering rate in the $\Lambda_{aa}$ scenario (in the $\Lambda_{aa'}$ scenario) given in Eq.~\eqref{Gammi} (in Eq.~\eqref{Gammiprime}). In the absence of scatterings and redshift, $g_i = 1$. 
  	 
Notice that since the scattering cross sections relevant for the neutrino evolution are independent of the energy (cf. Eqs.~\eqref{sigma} and~\eqref{sigmas}), $g_i$ also turn out to be independent of the energy, and therefore $g_{i}(z, E_\nu) \Longrightarrow g_i(z)$.
Since $\Gamma_i$ is proportional to $|U_{\tau i}|^2$, the functions $g_i$ are different for different mass states. In the SM this scattering is negligible and therefore we expect $F_{\nu_1} = F_{\nu_2} = F_{\nu_3}$, so we take a universal normalization for the three mass eigenstates $\mathcal{A}$.  Using Eq.~\eqref{relation}, one can write the neutrino fluxes on Earth as 
\be \label{FNUearth}
    {F}_{\nu_i} = {\mathcal{A}}\, E_\nu^{-\kappa}\, \exp\left[g_i(z =0)\right], 
\ee
where $\tilde{\mathcal{A}}=\mathcal{A} [r(0)]^2$ in which $r(0)$ is the present distance between the source and Earth (see Eq.~\eqref{r(z)}).

Neglecting the ALP regeneration term, we find the solution of  Eq.~\eqref{Fa0} as
\begin{align} \label{Faparamet}
    \tilde{F}_a(0, E_a) =& \int_0^{z_L} \left[ \int_{E_a}n_a \sum_{i,j} \frac{d\sigma_s(a+\nu_i(E_{\nu_i})\to a(E_a)+\nu_j)}{dE_a}\, \frac{\tilde{F}_{\nu_i}(z, E_\nu)}{H\, (1+z)}\, dE_{\nu}\right. \nonumber \\
    &+\left. \int_{E_a}\sum_{i,j} n_j \frac{d\sigma(\nu_j+\nu_i(E_{\nu_i})\to a(E_a)+a)}{dE_a}\, \frac {\tilde{F}_{\nu_i}(z, E_\nu)}{H\, (1+z)}\, dE_{\nu} \right] \exp[j(z, E_a)]\, dz\, ,
\end{align}
where the differential cross sections are given in Eqs.~\eqref{diffnunu} and~\eqref{diffsig}, and
\begin{equation}
    j(z, E_a) \equiv - \ln(1+z) - \kappa\, \left[1 - \frac{1}{1+z}\right] - \int^z_0 \frac{\Gamma_a}{H\, (1+z')}\, dz' \,,\label{j}
\end{equation}
in which $\Gamma_a$ accounts for the ALP scattering rate in the $\Lambda_{aa}$ scenario given by Eq.~\eqref{GammaALP}. Since $\Gamma_a$ is constant in energy, $j(z, E_a)$ is also independent of $E_a$. For the $\Lambda_{aa'}$ model, we can write similar relations, setting $n_a = 0$, substituting $d\sigma(\nu_i\nu_j \to aa)/dE_a$ with $d\sigma'(\nu_i\nu_j \to aa')/dE_a$ (see Eqs.~(\ref{sigma'})) and $\Gamma_a\to \Gamma_a'$ (see Eq.~\eqref{GammaALPrime}).

Before closing this section, let us discuss the validity of the approximation that we made by dropping the regeneration terms to obtain Eqs.~\eqref{g} and~\eqref{j}. Neglecting the neutrino regeneration terms leads to a slight underestimation of the neutrino flux arriving at Earth. In the extreme case where all neutrinos with energies greater than a nominal value $E$ are uniformly redistributed as neutrinos of lower energies, the change of flux at energies below $E$ will be $E^{- \kappa} \int (E^{-\kappa}/E) dE=1/\kappa$, which for $\kappa = 2$ is $\sim 50\%$. In reality, not all neutrinos above $E$ undergo scattering. For the parameter range that we are interested in only about 50\% of the neutrinos scatter off the background, so we expect the underestimation of the $\nu$ flux due to the neglect of the neutrino regeneration terms to be $\sim (20-30)$\%.  In Section~\ref{sec:NUfluxes},  we shall find that enhancing $g_i$ by 20\% to 30\%, the IceCube bound on $\mathcal{A}$ should be about 20\% to 30\% less stringent (see Eq.~\eqref{Abound}). Similarly, correcting for the ALP regeneration terms, we would obtain an estimate for $F_a$ about 20\% to 30\% higher, so our analytical formulas in Eqs.~\eqref{Faparamet} and~\eqref{j} give a conservative estimate. In computing the final photon flux,  the changes due to relaxation of the bound on $\mathcal{A}$ and due to the underestimation of $\mathcal{A}$ partially cancel each other out (rather than adding up). As a result, in the end, neglecting the regeneration terms would change the predicted photon flux only by about 10\% to 20\%. For the present general analysis, a correction of $\sim 10\%$ is irrelevant.
  	
\section{Bounds on the model \label{sec:bounds}}
In Section~\ref{sec:NUfluxes}, we review the bounds from IceCube on the neutrino flux from GRB 221009A accompanying the photon flux. We also discuss the implications of the discovery of TXS 056+0506 as a point source for high-energy neutrinos within our scenario. Then, in Section~\ref{sec:astro}, we comment on the impact of the new physics introduced in this paper on astrophysical and cosmological observations.

 \subsection{IceCube bounds on neutrino fluxes \label{sec:NUfluxes}}
The fluxes of active neutrinos from GRB  are the central elements of our proposal. Here, we summarize the relevant information on these fluxes. As explained in Appendix~\ref{app:Evolution}, it is enough and convenient to define $F_{\nu_i}$ as the sum of the $\nu_i$ and $\bar{\nu}_i$ fluxes to study their evolution and the production of ALPs along the path from the source to  Earth.

If the origin of the photon flux observed by LHAASO is high-energy neutrinos, we expect a large $\nu$ flux to arrive from GRB 221009A. IceCube has, indeed,  searched for a neutrino flux in a time window around the GRB 221009A~\cite{Abbasi:2023xhh}. The search has led to a null result, setting bounds for three distinct energy ranges: $i)$ $[0.5,\, 5]$~GeV; $ii)$ a few GeV-TeV, and $iii)$ TeV-PeV, assuming a power law spectrum with $\kappa = 1.5$, 2, 2.5, and 3. Among these bounds, the strongest (and most relevant for our scenarios) is the third energy range. We discuss it below, taking $\kappa=2$.  In our analysis, we also examined other values of $\kappa$. We will comment on the robustness of our result against variations of $\kappa$ in Section~\ref{sec:LHAASO}.

IceCube provides a limit on the integrated flux of muon neutrinos over a time period $t_\text{IC} = 3$ hours around the arrival time of the 18~TeV event~\cite{Abbasi:2023xhh}:
\be \label{ICECUBE}
    {F}_{\nu_\mu}^\text{int} = \int_{t_\text{IC}} dt { F}_{\nu_\mu} (t)  < \frac{3\times 10^{-2}}{E_\nu^2}~ \frac{\rm GeV}{\rm cm^2}\,,
\ee
where ${F}_{\nu_\mu} \equiv \sum_i |U_{\mu i}|^2\, {F}_{\nu_i}$. Then the flux itself is restricted as
\be \label{eq:fmubound}
    {F}_{\nu_\mu} = \frac{F_{\nu_\mu}^{int}}{t_\text{IC}}  <   2.8 \times 10^{-6} E_\nu^{-2}\, {\rm cm}^{-2} {\rm sec}^{-1} {\rm GeV}.
\ee
Using Eq.~\eqref{FNUparamet}, the IceCube bound~\eqref{eq:fmubound} for $0.1~{\rm TeV} < E_\nu < 1000$~TeV and  $\kappa=2$ implies 
\be \label{Abound}
    \mathcal{A} < \left( \sum_i |U_{\mu i}|^2\exp[ g_i(z=0)]\right)^{-1}\times \frac{2.78 \times 10^{-6}~{\rm GeV}}{{\rm cm}^2\, {\rm sec}}\,.
\ee
As mentioned above, IceCube also reports weaker bounds on neutrino flux at lower energies~\cite{Abbasi:2023xhh}. We have checked and confirmed that for a power law spectrum, these low-energy bounds are automatically satisfied once the TeV range bound in Eq.~\eqref{ICECUBE} is respected.

Since its inception, IceCube has collected a significant number of high-energy neutrinos with $E_\nu > 100$~TeV. 
Two point sources for such neutrinos have been identified. One of them,  TXS 0506+056, is located at a redshift of 0.33. If $\Gamma_i$ (or $\Gamma_i^\prime$ ) is too large, we expect a very high suppression of the flux from TXS 0506+056. We set a limit on the optical depth at $z=0.33$, $\tau_{\nu_i}(z=0.33)$, to be smaller than 5 (corresponding to suppression of $\exp(-5) \simeq 0.67\%$):
\be \label{optical-depth}
    \tau_{\nu_i}^{(\prime)}(z=0.33)\equiv \int_0^{z=0.33} \frac{\Gamma_i^{(')}}{H(1+z)}\, dz < 5\,.
    \ee
For $\nu_1$ ($\nu_3$), this  implies $\Lambda_{aa}> 465$~MeV ($\Lambda_{aa}> 850$~MeV)  and $\Lambda_{aa'}>405$~MeV ($\Lambda_{aa'}>740$~MeV).

It is now established that about 10\% of the total diffuse high-energy cosmic neutrinos observed by IceCube come from our own galaxy~\cite{IceCube:2023ame}. Across the galactic distance, the suppression of neutrino flux is negligible because the mean free path of neutrinos is much larger. The origin of the rest of diffuse neutrinos detected by IceCube is unknown, but within our model, neutrino sources should be at redshifts below about 0.3; otherwise, the neutrino flux would be strongly attenuated.

\subsection{Astrophysical and cosmological bounds\label{sec:astro}}
For the ALP mass range of interest, the strongest bound on $g_{a\gamma \gamma}$ comes from magnetic white dwarf polarization~\cite{Dessert:2022yqq}:
\begin{equation}
    |g_{a \gamma \gamma}| < 5.4\times 10^{-12}\, {\rm GeV}^{-1}\,.
\end{equation}
There are also more stringent bounds for $m_a<\mathcal{O}(1)$~neV from the Planck observations, but these bounds are based on the assumption that the whole dark matter is composed of axion~\cite{Escudero:2023vgv} so do not apply to our scenarios.

As discussed in Section~\ref{sec:NuALP}, the interactions in Eqs.~\eqref{aanunu}  or~\eqref{aaprimenunu} can lead to ALP production in the early universe before neutrino decoupling. ALPs can reach thermal equilibrium before the BBN era provided that the rate of ALP production per neutrino (that is, $\sigma(\nu_\tau\nu_\tau \to aa)\, n_{\nu_\tau} \sim (56~{\rm cm}^{-3}(T/T_0)^3) / (32\pi\Lambda_{aa}^2)$ in which $T_0 \simeq 1.9$~K is the neutrino temperature today) is larger than the Hubble expansion rate at the time of neutrino decoupling. As long as $\Lambda_{aa}~{\rm or}~\Lambda_{aa'} < 5\times 10^5$~GeV, this condition is fulfilled and ALPs can reach thermal equilibrium in the early Universe, contributing to extra-relativistic degrees of freedom $\Delta N_\text{eff} =  4/7$.  This contribution is already ruled out by the BBN~\cite{Mangano:2011ar} and the CMB data~\cite{Planck:2018vyg}. Another important bound comes from the free streaming of neutrinos at redshifts between 2000 and $10^5$~\cite{Taule:2022jrz}. Unless $\Lambda_{aa}>100$~TeV, the process $\nu+\nu\to a+a$ can prevent free streaming at $z\sim 10^5$; that is, for $\Lambda_{aa}<100$~TeV, $\sigma(\nu \nu\to a a)\, n_\nu|_{z=10^5} > H(z=10^5)$. The $10^5$ redshift corresponds to a temperature of the Universe about 20~eV. If $m_{a'}>50$~eV, the processes $\nu \nu\to aa'$ do not take place at $z < 10^5$, and no bound can be set on $\Lambda_{aa'}$ from the requirement of free streaming. In Appendix~\ref{app:UVcompletion} where the UV completion of the models is introduced, we also show that the cosmological bounds can be avoided by invoking varying mass for $a'$ or for the mediator of the interaction in Eq.~\eqref{aanunu}. Notice that in our model,  by construction, the contribution from ALP to the energy budget of the Universe is negligible,  and consequently, ALPs cannot account for dark matter.
 
Additionally, ALPs can be produced in supernova cores via couplings to neutrinos.  In the protoneutron star, ALPs scatter off neutrinos and can be trapped inside the supernova core. They can contribute to the supernova cooling by escaping from the neutrinosphere. 
The measurement of the total energy of neutrinos emitted from SN1987a suffers from an uncertainty of a factor of 2. On the other hand, the prediction of the binding energy is also subject to a similarly large uncertainty~\cite{Kamiokande-II:1987idp, Bionta:1987qt}. As a result, if, along with neutrinos, new particles diffuse out of the neutrinosphere, the energy loss due to them can hide in the uncertainties, as long as their total luminosity is not much larger than the total luminosity of neutrinos and antineutrinos.  Each scalar degree of freedom such as $a$ or $a'$, coming to thermal equilibrium with $\nu_\tau$ will have a luminosity of
\begin{equation}
    \frac14\, \frac{\int \frac{E^3}{e^{E/T} - 1}\,dE}{\int \frac{E^3}{e^{E/T} + 1}\, dE}=0.28
\end{equation}
relative to the luminosity of $\nu_\mu +\bar{\nu}_\mu+\nu_\tau+\bar{\nu}_\tau$. Thus, the emission of $a$ and $a'$ from $\nu_\tau$ neutrinosphere can be tolerated within the present uncertainty.
In fact, $a'$  decays back to pairs of tau neutrinos and ALP. Thus, on average 2/3 of the energy of $a'$ will arrive at Earth in the form of neutrinos, so the contribution of $a'$ to the missing energy is further reduced.

Let us now discuss the possible impacts of $a$ and $a'$ on the formation of structures. As long as there is a mechanism to prevent the production of new light particles before neutrino decoupling, the timing of matter-radiation equality will remain intact even if the relativistic particles convert to each other. As a result, the onset of efficient structure formation will be similar to that in the SM. As discussed before, in the $\nu\nu \to aa$ model,  the background $\nu_3$ and $\nu_2$ will be converted to lighter $a$ and $\nu_1$ particles, so they do not contribute to the dark matter content. As long as the sum of the neutrino masses is less than the limit of 0.12~eV~\cite{Planck:2018vyg}, the effect will not be discernible because the contribution of neutrinos to the matter budget of the universe is too small~\cite{Lesgourgues:2012uu}.  The contribution of $\nu_2$ and $\nu_3$ to dark matter within SM is less than 1\% so their total disappearance cannot significantly affect structure formation.

\section{\boldmath High energy $\gamma$ fluxes from GRB 221009A \label{sec:GRB}}
In Section~\ref{sec:LHAASO}, we show that within our scenarios the LHAASO 18~TeV event can be explained. We then discuss the photon-neutrino flux relation characteristic of our solution. In Section~\ref{sec:Carpet}, we comment on the possibility to explain the 251~TeV photon event observed by Carpet-2 within our scenarios.
 
\subsection{LHAASO \label{sec:LHAASO}}
Taking an effective area of LHAASO $A_\text{eff}^\text{LHAASO}=0.5$~km$^2$  and a data taking time $t_\text{LHAASO}=2000$~sec (cf. Eq.~(4) of Ref.~\cite{Finke:2022swf}), we find that observation of  a single event with $E_\gamma=18$~TeV sets a 95\%~C.L. lower bound on the photon flux:
\be
    F_\gamma (E_\gamma = 18~{\rm TeV}) > 2.9 \times 10^{-19}~{\rm cm}^{-2}\, {\rm sec}^{-1}\, {\rm GeV}^{-1}.
    \label{LOW-LHAASO}
\ee

Let us discuss the predictions of our scenarios for the photon flux. In the galactic magnetic field, ALPs are converted to photons with the same energy. Thus, the photon flux, $F_\gamma$, can be written as 
\be
    F_\gamma (E_\gamma) = {F}_a(z=0,E_a)\, P_{a \to \gamma}\,, 
\ee 
where $E_\gamma=E_a$ and $P_{a \to \gamma}$ is the conversion probability given in Eq.~\eqref{Pagamma}. In Appendix~\ref{app:ALPconversion}, more information about this probability can be found. We compute the photon flux using the relations in Section~\ref{sec:ALPsNUs}, assuming that neutrino fluxes saturate the IceCube bound~\eqref{ICECUBE} and~\eqref{Abound}.
For the conversion probability, we use the results found in Ref.~\cite{Carenza:2022kjt} which invoke the galactic magnetic field model presented in Ref.~\cite{Jansson:2012pc}. According to Ref.~\cite{Carenza:2022kjt}, the uncertainty in the $a \to \gamma$ conversion probability in our galaxy for ALPs coming from the direction of GRB221009A is $\sim 10\%$.

\begin{figure}[t!]
    \def\sepf{0.54}
	\centering
    \includegraphics[scale=\sepf]{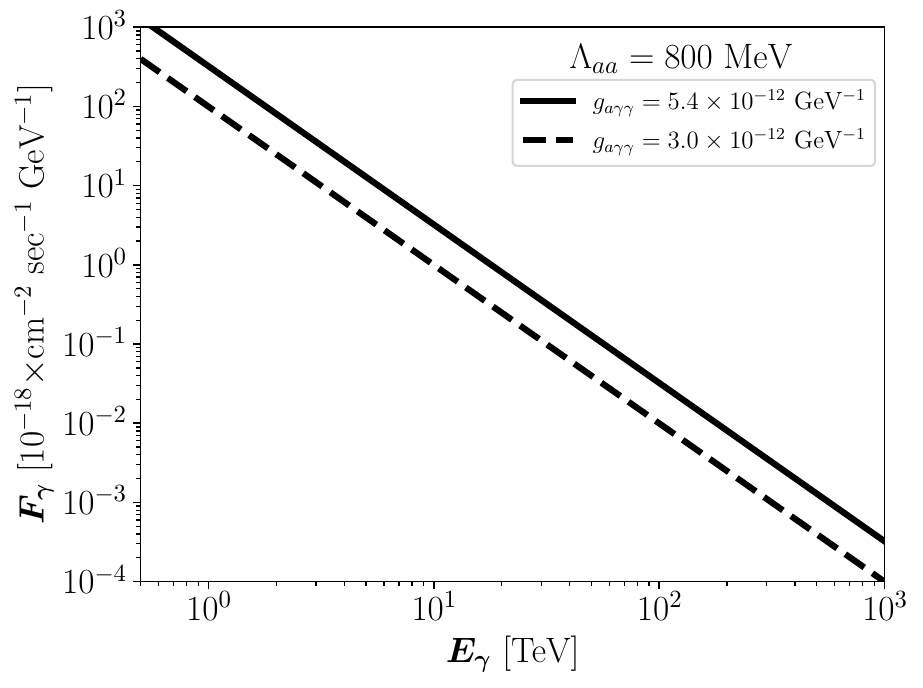}
    \includegraphics[scale=\sepf]{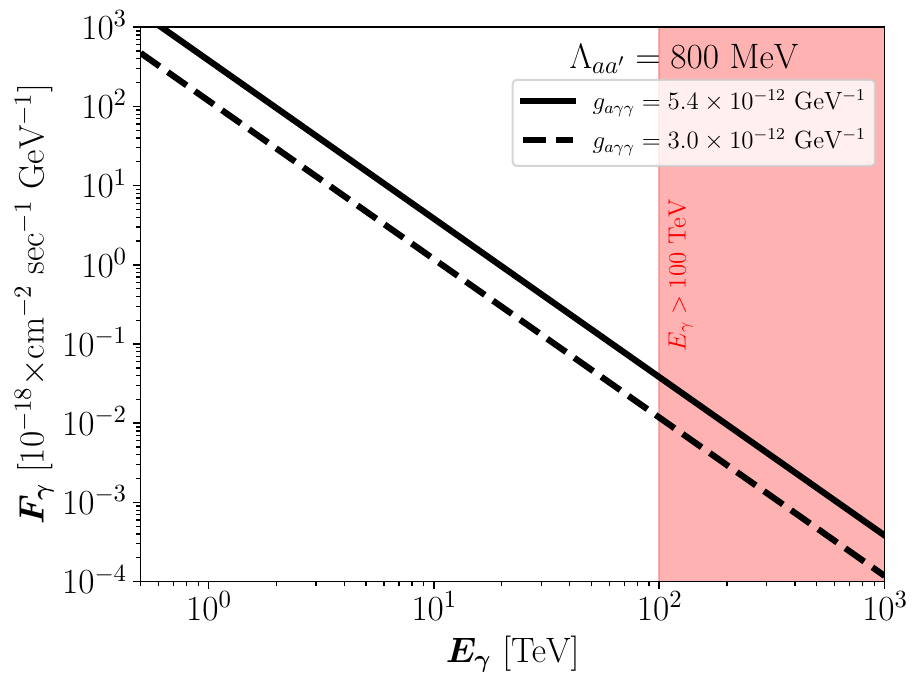}
    \caption{Predicted $\gamma$-ray flux $F_\gamma$ as a function of the photon energy $E_\gamma$, for $\Lambda_{aa} = 800$~MeV (left) or $\Lambda_{aa'} = 800$~MeV (right), and $g_{a\gamma\gamma}= 5.4 \times 10^{-12}$~GeV$^{-1}$ or $g_{a\gamma\gamma}= 3.0 \times 10^{-12}$~GeV$^{-1}$ (corresponding respectively to $P_{a\to\gamma} \simeq 4.2 \times 10^{-4}$ or $1.5 \times 10^{-4}$). In the right panel, the red region represents the region where the validity of the EFT is not guaranteed.}
    \label{fig:spectrum}
\end{figure} 
Figure~\ref{fig:spectrum} shows the shape of spectrum taking $\Lambda_{aa}, \ \Lambda_{aa'}=800$~MeV and assuming constant $P_{a \to \gamma}$. As discussed in Appendix \ref{app:ALPconversion}, for the energy range shown in Fig.~\ref{fig:spectrum}, the variation in $P_{a \to \gamma}$ is indeed negligible; however, for energies below TeV, $P_{a \to \gamma}$  decreases, suppressing $F_\gamma$. Thus, as discussed in Appendix~\ref{app:ALPconversion}, the bounds on the lower-energy photon flux can be avoided. As seen in the figure, the shape of $F_\gamma$ also follows a power law that can be parametrized as $F_\gamma(E_\gamma) \propto E_\gamma^{-\kappa}$.
The reason is that with the effective interactions in Eqs.~\eqref{aanunu} and~\eqref{aaprimenunu} the cross section is independent of the energy. The shaded range in the right panel indicates a tentative limit above which the energy-momentum transfer in the scattering $\nu(E_\nu)\, \nu_3(m_3)\to a\, a$ becomes comparable to the mass of the mediator, so we cannot use the effective interaction coupling in Eq.~\eqref{aaprimenunu}. In the right panel, we do not need to worry about the validity of $\Lambda_{aa}$ in this energy range, as there is no background of $\nu_3$ or $\nu_2$, and the center of mass energy in the scatterings off background will be low enough to guarantee the validity of the effective interaction. Notice that to draw these figures, we have neglected the regeneration terms. Taking into account their effects may cause a slight deviation from the power law with $\kappa=2$ for $F_\gamma$ even with the scattering cross sections being constant in energy.
  
\begin{figure}[t!]
    \def\sepf{0.54}
	\centering
    \includegraphics[scale=\sepf]{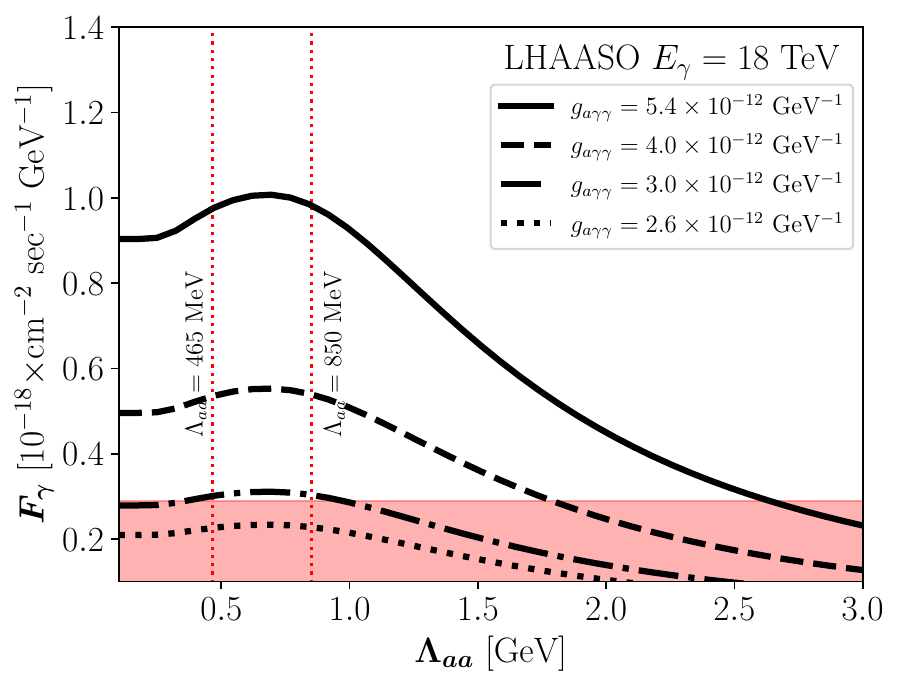}
    \includegraphics[scale=\sepf]{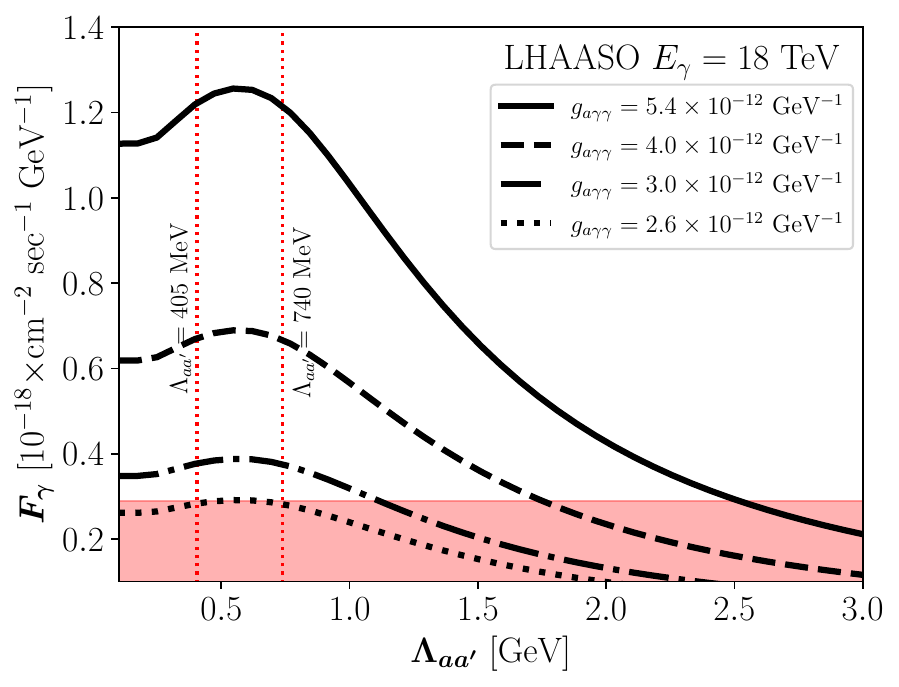}
    \caption{The predicted $\gamma$-ray flux $F_\gamma$ as a function of $\Lambda_{aa}$ (left) and $\Lambda_{aa'}$ (right), for $E_\gamma = 18$~TeV and $g_{a\gamma\gamma}= (5.4,\, 4.0,\, 3.0,\,  2.6) \times 10^{-12}$~GeV$^{-1}$.
    In the horizontal red region, the photon flux is too small to explain the LHAASO 18 TeV photon observation; see Eq.~\eqref{LOW-LHAASO}. The vertical dotted lines show limits below which the high-energy $\nu_1$ and $\nu_3$ fluxes from TXS 0506+056 are severely suppressed by scattering off relic backgrounds.}
    \label{fig:Flux}
\end{figure} 
Figure~\ref{fig:Flux} shows the predicted photon fluxes at  $E_\gamma = 18$~TeV versus $\Lambda_{aa}$ (left) or $\Lambda_{aa'}$ (right), for four values for the coupling $g_{a\gamma\gamma}= (5.4,\, 4.0,\, 3.0,\, 2.6) \times 10^{-12}$~GeV$^{-1}$ corresponding to $P_{a\to \gamma} = (1.1,\, 1.5,\, 2.7,\, 4.9) \times 10^{-4}$; see Eq.~\eqref{Pagamma}. The vertical dotted lines show the limits below which the high-energy $\nu_1$ and $\nu_3$ fluxes from TXS 0506+056 are severely suppressed by scattering off relic backgrounds. Thus, the left parts of the plots below $\Lambda_{aa}=465$~MeV and $\Lambda_{aa'}=405$~MeV are ruled out by the observation of neutrinos from TXS 0506+056 (see Section~\ref{sec:NUfluxes} for more details). In the shaded light red region, the 18~TeV photon flux is too small to explain the LHAASO event. As seen in the figures taking the maximum value of $g_{a\gamma \gamma}$, the predicted flux for $\Lambda_{aa^{(\prime)}}<2500$~MeV lies above this lower bound, so that the models with and without $a'$ can explain the 18~TeV photon event. From this figure we find that to explain the LHAASO observation, $g_{a\gamma\gamma}$ must be larger than about $3 \times 10^{-12}$ GeV$^{-1}$. Such ALP photon coupling values can be tested by forthcoming measurements~\cite{Escudero:2023vgv}. The curves have a peak around 700$-$800~MeV. Above the peak, the flux decreases as a result of the reduction in ALP production. Below 700~MeV, the ALP scattering decreases the final photon flux. With $a'$, the flux can be higher than the maximum amount without $a'$. This is because, without $a'$, for $\Lambda_{aa}<800$~MeV scattering of $a$ off-background ALPs can severely suppress the final photon flux. What limits the predicted photon flux is, of course, the bound on the neutrino flux from GRB 221009A by IceCube. If this bound was stronger by a factor of about four, our solution for the 18~TeV LHAASO event would not be viable. It is tantalizing to add a coupling of {\it e.g.} ${a'}^2\nu^T\, c\, \nu$ to further suppress the neutrino flux reaching Earth and avoid the IceCube bound; however, such suppression would imply a greater suppression for neutrinos from TXS 0506+056 at $z=0.33$ which defies observation. We have checked our results against the variations of $\kappa$ and have found them to be robust. 

\begin{figure}[t!]
    \def\sepf{0.54}
	\centering
    \includegraphics[scale=\sepf]{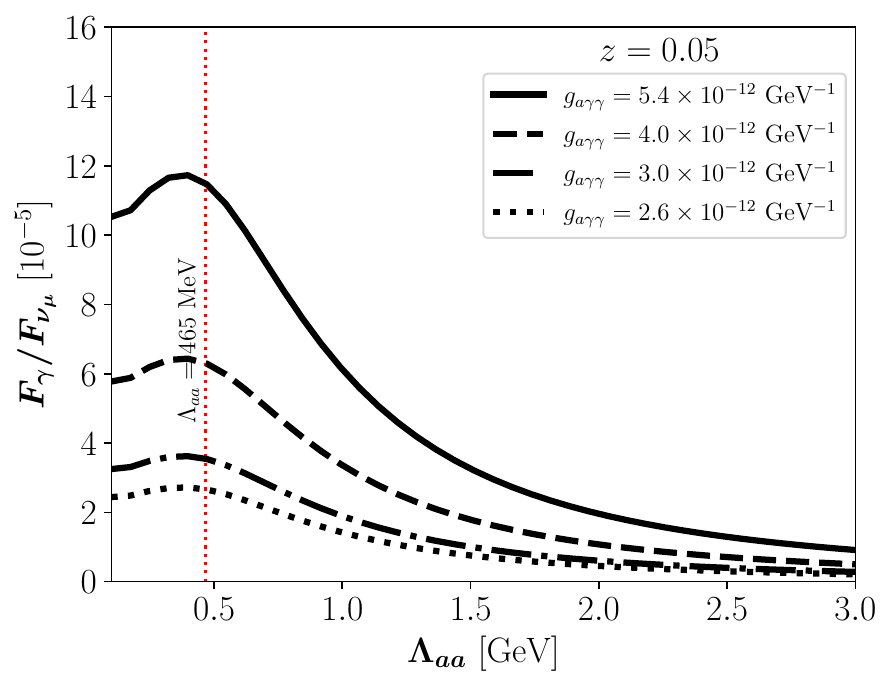}
    \includegraphics[scale=\sepf]{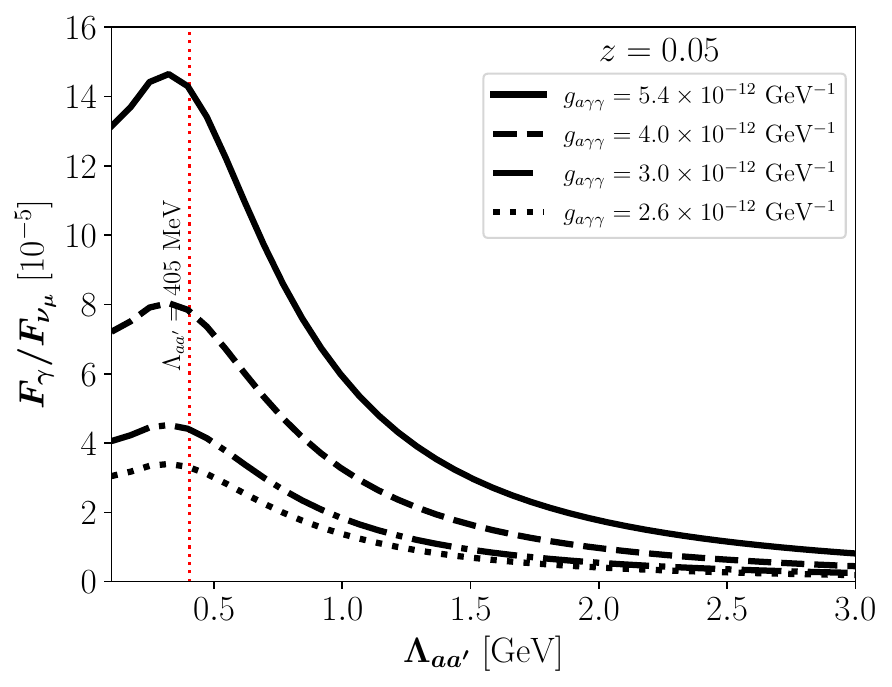}
    \includegraphics[scale=\sepf]{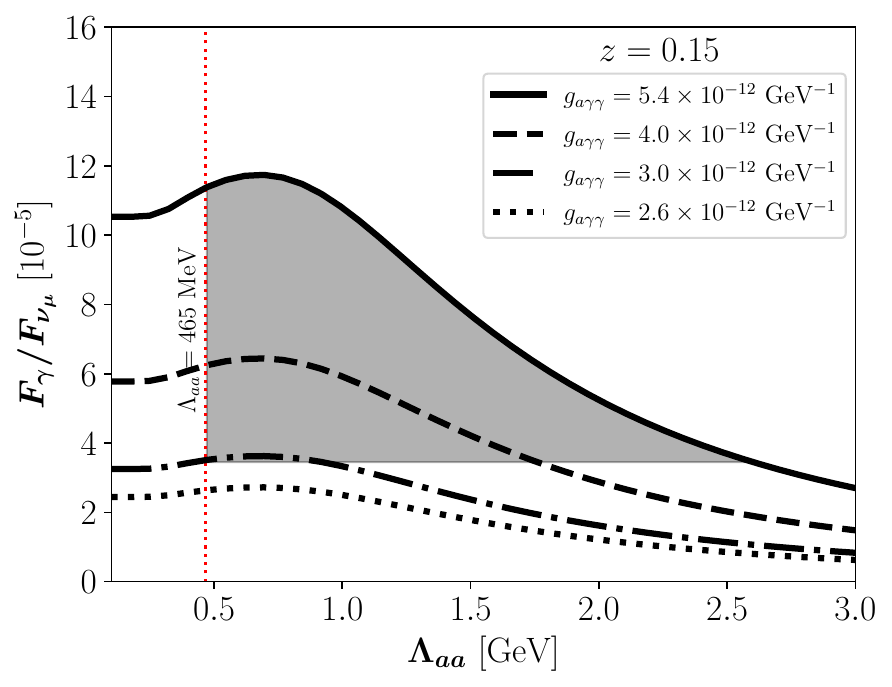}
    \includegraphics[scale=\sepf]{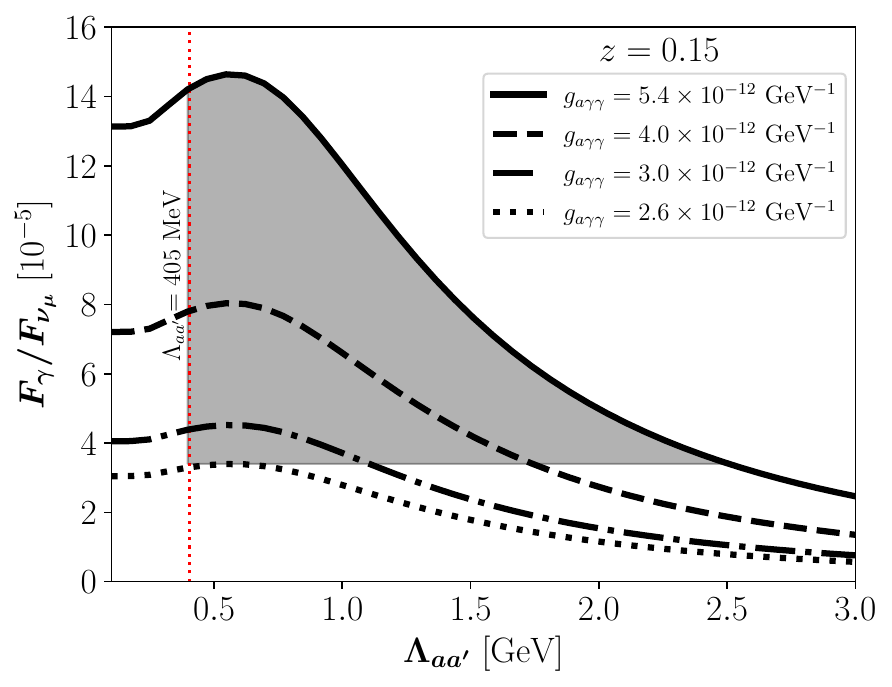}
    \caption{Ratio of fluxes $F_\gamma/F_{\nu_\mu}$ as a function of $\Lambda_{aa}$ (left) and $\Lambda_{aa'}$ (right), assuming a source at $z = 0.05$ (top) and $z = 0.15$ (bottom) and  taking different values of the coupling $g_{a\gamma\gamma}$. 
    The gray regions in the lower panels provide an explanation for the LHAASO observation without violating any existing bounds.}
    \label{fig:FluxRatio}
\end{figure} 
As we discussed above, $F_\gamma/F_{\nu_\mu}$ is almost constant in energy in the range of our interest ($1~{\rm TeV} \lesssim E \lesssim 100~{\rm TeV}$). However, the ratio depends on $\Lambda_{aa}$ or $\Lambda_{aa'}$, as well as on the redshift of the source. Figure~\ref{fig:FluxRatio} shows the ratio $F_\gamma/F_{\nu_\mu}$ for sources at redshifts $z =  0.05$ (top) and 0.15 (bottom) as a function of the scales $\Lambda_{aa}$ (left) or $\Lambda_{aa'}$ (right) for various values of the ALP-photon coupling (recall that $F_{\nu_\mu} \equiv \sum_i |U_{\mu i}|^2 F_{\nu_i}$). Although $F_a/F_{\nu_\mu}$ is of the order $0.1-1$, $F_\gamma/F_{\nu_\mu}$ turns out to be of the order $(1-10) \times 10^{-5}$ because $P_{a\to \gamma}\sim 10^{-4}$. Higher redshifts give rise to a peak at higher values of $\Lambda_{a a^{(\prime )}}$. This is understandable as the distance from the source increases and so does the absorption probability of ALPs. Note that these predictions are independent of the details of the source. In fact, the source can be any neutrino engine (AGN, GRB, etc.), either transient or stationary. If future multi-messenger searches find $F_\gamma/F_{\nu_\mu}$ above the curve corresponding to $g_{a\gamma \gamma}=5.4\times 10^{-12}$ GeV$^{-1}$ (or any other bound on $g_{a\gamma \gamma}$ achieved by then), our solution will be ruled out. The gray region in the panel $z=0.15$ provides an explanation for the LHAASO observation. As shown by the lower boundary of this gray region, to provide a solution for the LHAASO observation, for both scenarios and independently of the value of $g_{a \gamma\gamma}$, the ratio of the fluxes should be larger than $\sim 3.3\times  10^{-5}$, which is the ratio of the bounds in Eqs.~\eqref{eq:fmubound} and~\eqref{LOW-LHAASO} at an energy of 18~TeV.

As shown in Ref.~\cite{Murase:2015xka}, there is a weak tension between the IceCube neutrino flux with energy above 10~TeV and the gamma-ray flux in the energy range of 3 GeV to 1~TeV observed by Fermi-LAT. The basis of the tension is as follows: It is expected that the hadronic processes that produce charged pions and therefore neutrinos also produce $\pi^0$ with a ratio of $O(1)$, which decay to $\gamma \gamma$ with similar energy cascading down to lower energies. However, the tension is removed if the sources are opaque to the photons. At first sight, it seems that our prediction of photons emerging from a new neutrino interaction will make this tension more severe.
However, the photon flux that we predict is irrelevant for this tension because $i)$ the photons accompanying neutrinos with energy above few TeV have energies of the same order and do not contribute to the photon flux with energy below TeV discussed in Ref.~\cite{Murase:2015xka}. For lower energies, as discussed in Appendix~\ref{app:ALPconversion}, the ALP-photon conversion is not effective. $ii)$ From Fig.~\ref{fig:FluxRatio}, we observe that $F_\gamma/F_\nu < 2 \times 10^{-4}$ but the tension pointed out in Ref.~\cite{Murase:2015xka} deals with $F_\gamma \sim F_\nu$  so this small correction is irrelevant.

\subsection{Carpet-2\label{sec:Carpet}}
If the 251~TeV event registered by Carpet-2 indeed originates from a high-energy photon (rather than from various background sources), we obtain a lower bound on the flux of $\sim 100$~TeV photons. Taking $A_\text{eff}^\text{Carpet-2} = 25$~m$^2$ and $t_\text{Carpet-2} = 4536$~sec, the following bound can be set at the 95\%~CL~\cite{Finke:2022swf}:
\be
    \label{CARPET}
    F_\gamma(E_\gamma = 251~{\rm TeV}) > 1.8 \times 10^{-16}~{\rm cm}^{-2}\, {\rm sec}^{-1}\, {\rm GeV}^{-1}.
\ee
Notice that if such a large flux reached Earth during the 3-hour data taking of LHAASO, we would expect $\sim 5000$ events at LHAASO with an energy of a few 100~TeV.  Null results from LHAASO could be explained by assuming that this flux of a few 100~TeV energy arrives at Earth only after LHAASO ceases to monitor GRB 221009A.  

Let us check whether the Carpet-2 signal can be explained within our scenario. We assume that all neutrinos with an energy $E_\nu$ higher than $E_\text{Carpet-2} = 251$~TeV are converted to $a$ with uniformly distributed energies in the range $[0,\, E_\nu]$. Subsequently, the produced ALPs convert into photons with a probability  
given in Eq.~\eqref{Pagamma}. Then, the flux of photons at 251~TeV equals  
\begin{equation} \label{eq:gcarp}
    F_\gamma^\text{Carpet-2} =  P_{a \rightarrow \gamma} N_{\gamma/\nu}\int_{> E_\text{Carpet-2}} \frac{F_\nu^\text{SM}(E_\nu)}{E_\nu}\, dE_\nu\,,
\end{equation}
where $F_\nu^\text{SM}(E_\nu)$ is the neutrino flux at Earth in the absence of absorption and $N_{\gamma/\nu}$ is the number of $\gamma$ per neutrino. Parameterizing the neutrino flux as  
\begin{equation}
    F_\nu^\text{SM}(E_\nu) = \mathcal{B}\, E_\nu^{-\kappa}     
\end{equation}
we obtain from Eq.~\eqref{eq:gcarp}
\begin{equation}
    F_\gamma^\text{Carpet-2} = \mathcal{B}\,  P_{a \rightarrow \gamma} \frac{N_{\gamma/\nu}}{\kappa}  E_\text{Carpet-2}^{-\kappa},
\end{equation}
and consequently, 
\be
    \mathcal{B}\, =    F_\gamma^\text{Carpet-2}\frac{\kappa}{N_{\gamma/\nu} P_{a \rightarrow \gamma}}E_\text{Carpet-2}^{\kappa}. 
\ee
For $N_{\gamma/\nu}= 2$, $\kappa = 2$ and $F_\gamma^\text{Carpet-2}$ satisfying bound Eq.~\eqref{CARPET}  we obtain 
\begin{equation}
    \mathcal{B} \simeq 0.075~{\rm GeV~cm}^{-2}~{\rm sec}^{-1}\frac{1.5\times 10^{-4}}{P_{a \to \gamma}}. 
\end{equation}
Comparing this with the IceCube bound Eq.~\eqref{eq:fmubound} for $\kappa = 2$, we find that the neutrino flux should be suppressed by the factor $e^{-\tau_\nu} \sim 4\times10^{-5}$ which corresponds to an optical depth of 10 at redshift of 0.15: $\tau_\nu(z=0.15)=10$.

The expected neutrino flux from farther sources is even further suppressed. In particular, the neutrino flux from the blazar TXS 0506+056 identified by IceCube at a redshift $z = 0.33$ would be suppressed by a factor $\exp[- \tau_\nu (0.33)]$. The optical depth can be estimated as
\begin{equation}
    \tau_\nu (0.33) = \tau_\nu (0.15) \times\frac{\int_0^{0.33}(1+z)^2/H dz}{\int_0^{0.15}(1+z)^2/H dz} \simeq 25\,,  
\end{equation}
which gives suppression factor $\sim 10^{-11}$. With such strong suppression, observation of events from TXS 0506+056 requires an unreasonably powerful source. In other words, the explanation of the Carpet-2 result within our scenarios implies a huge suppression of the neutrino flux via $\nu\, \nu\to a\, a$, which contradicts observations of neutrinos from an even more distant source TXS 0506+056. Making non-observation of the $\nu$ flux by IceCube from GRB 221009A compatible with the Carpet-2 event and the observation of $\nu$ flux from far-away sources requires something beyond these scenarios. In fact, the findings of the HAWC collaboration indicate a galactic origin of the 251 TeV Carpet-2 event~\cite{2022ATel15675....1F}. Therefore, we assume that the Carpet-2 event is only a background.

\section{Summary and discussions\label{sec:Conclusions}}
Motivated by the observation of the anomalously high-energy photon event from GRB 221009A, we propose scenarios that predict a photon flux from powerful neutrino sources such as Gamma Ray Bursters (GRB) or Active  Galactic Nuclei (AGN) at cosmological distances, with energies even exceeding the limit above which the Universe becomes opaque for the photons. The proposal involves an ALP coupled to the photon through the interaction in Eq.~\eqref{aFF}. ALPs are produced via the scattering of high-energy neutrinos from the source off neutrinos of the cosmic background C$\nu$B. Then, after propagating cosmic distances, the high-energy ALPs reach our galaxy and convert to photons in the galactic magnetic field. In order for the ALP photon conversion to be efficient, the ALP mass should be in the $[0.01,\, 100]$~neV range.

We have proposed two scenarios: $i)$ A minimal setup with an effective coupling of the form given in Eq.~\eqref{aanunu}. In this scenario, the C$\nu$B is composed of only the lightest neutrino mass eigenstate because the lifetimes of the heavier neutrino mass eigenstates are shorter than the age of the Universe. A background of relic ALPs is unavoidable in this scenario. $ii)$ The scenario including a light scalar $a'$ with a mass in the range $\sim$ [50~eV, 1~keV] and an effective coupling given in Eq.~\eqref{aaprimenunu}. In this case, all three neutrino species have a lifetime larger than the age of the Universe so the C$\nu$B includes them all. The ALP background can be suppressed. In both scenarios, the high-energy ALP produced by couplings to neutrinos can scatter off the background relic neutrinos via the same interaction. In the first case, they can also scatter off the background ALPs. These scatterings suppress the flux of high-energy ALPs and consequently reduce the high-energy photon flux. As a result, the final photon flux as a function of the coupling constant $\Lambda$ has a peak, as shown in Fig.~\ref{fig:Flux}. That is, the $\gamma$ flux cannot be arbitrarily increased by enhancing the ALP neutrino interaction.

In our scenarios, the high-energy photon flux is produced and accompanied by the high-energy neutrinos. The values of the flux and the energy spectra of the high-energy photon and neutrino fluxes reaching Earth are correlated, as shown in Fig.~\ref{fig:FluxRatio}. This provides a means to test this solution for the GRB 221009A puzzle because, in the alternative solutions suggested in the literature, such a correlation does not exist.  The null result from IceCube searches for high-energy neutrinos from GRB 221009A sets an upper bound on the predicted high-energy $\gamma$ flux. Despite this bound, there are ranges of couplings for which the 18~TeV photon signal detected by LHAASO can be explained (cf. Fig.~\ref{fig:Flux}). The high-energy neutrino and photon fluxes are also accompanied by a concurrent high-energy ALP flux; however, its direct detection is very challenging even with next-generation haloscopes. On the contrary, the entire range of ALP-photon coupling consistent with the 18~TeV LHAASO photon ($g_{a\gamma\gamma} \gtrsim 2 \times 10^{-12}$~GeV$^{-1}$) could be tested by improvement on the bound by a factor of $\sim 2$ in the near future.

We have studied the possibility to explain the 251~TeV event reported by Carpet-2 within our scenario.  Explaining this event requires a huge flux of neutrinos from the source. To satisfy the IceCube bound on neutrinos from GRB 221009A requires an optical depth of 10 at a redshift of 0.15. Then, the optical depth for neutrinos from TXS 0506+056 at $z=0.33$ would be 25, which would defy the observation of high-energy neutrino flux from this point source. However, we should bear in mind that the Carpet-2  event can originate from a background. In fact, the HAWC collaboration has found evidence in favor of its galactic origin~\cite{2022ATel15675....1F}

For completeness, we have built UV-complete models that accommodate the particles and interactions of the suggested scenarios. These models involve new light particles at the MeV scale, which can be produced in the core-collapse supernova or in the meson or lepton decays. We show that bounds from these observations can be satisfied; however, improving the uncertainties in the supernova evolution and/or terrestrial meson or lepton decays can test these models. Throughout this paper, our focus was on the couplings of ALPs to tau neutrinos because new couplings of $\nu_\tau$ are less constrained than those of $\nu_\mu$ or $\nu_e$.

In our scenarios, the contribution from relic ALPs to the energy budget of the Universe is too low to account for dark matter. We show that if dark matter is composed of ultralight scalars coupled to the new particles introduced in our models, the production of these new particles in the early Universe can be prevented, because of obtaining varying masses. Therefore,  the bounds on extra relativistic degrees of freedom from BBN and CMB are respected. Moreover, this mechanism guarantees the free streaming of neutrinos in the recombination era, as required by the CMB data.

Although we have focused on the LHAASO event in this paper, our proposal may have wider applications. In general, we introduced neutrino-ALP interactions and studied their astrophysical and cosmological consequences. We derived evolution equations for a system of neutrinos and ALPs which can be applied for studies of propagation of high-energy neutrinos from various sources, generation of ALP, and $\gamma$ fluxes.

\subsection*{Acknowledgments}
The authors are grateful to Thomas Schwetz and M. M. Sheikh-Jabbari for useful comments and encouragement.
YF has received financial support from Saramadan under contract No. ISEF/M/401439. She would like to acknowledge support from ICTP through the Associates Programme and from the Simons Foundation through grant number 284558FY19 as well as under the Marie Sklodowska-Curie Staff Exchange grant agreement No 101086085-ASYMMETRY.
NB received funding from the Spanish FEDER / MCIU-AEI under the grant FPA2017-84543-P. NB and YF thank the ICTP staff for their hospitality during their stay when this study was completed. This work has been supported by the European Union's Framework Programme for Research and Innovation Horizon 2020 under grant H2020-MSCA-ITN-2019/860881-HIDDeN.

\appendix
\section{Evolution equations for neutrino and ALP fluxes \label{app:Evolution}} 
In this appendix, we derive the evolution of the fluxes $F_{\nu_i}$ for the sum of neutrinos and antineutrinos,  and $F_a$ for the flux of ALPs, from the source to  Earth. Let us first neglect the interactions and derive the relevant formulas for a noninteracting flux propagating in the metric of $\Lambda$CDM
\begin{equation}
    d^2s = d^2t - a^2(t)\, \delta_{ij}\, dx^i\, dx^j,
\end{equation}
where $a = 1/(1+z)$ is the scale factor, $z$ the redshift, and the Hubble expansion rate $H$
\be
    H \equiv \frac{\dot{a}}{a} = H_0 \left[\Omega_\Lambda+(1-\Omega_\Lambda)\, (1+z)^3\right]^{1/2},
\ee
where $dt = -dz/(H(1+z))$. The distance between the source at redshift $z=0.15$ and a redshift of $z$ can be written as
\be \label{r(z)}
    r(z) = \int dx = \int \frac{dt}{a(t)} = \int_z^{0.15} \frac{dz}{H}\,.
\ee
The areal distance is given by $a\times r$, the distance to our galaxy corresponding to $r(0)$. Let us introduce the comoving distribution of the flux $F^0(z, E_0)$ where $E_0 \equiv E(z)/(1+z)$. Thus,
\be \label{FF0}
    F(z, E(z))\, dE = F^0(z, E_0)\, dE_0 \qquad \rightarrow \qquad F(z, E(z)) = F^0(z, E_0)/(1+z)\,.
\ee
As long as the number of particles of a given species is conserved, $F^0(z, E_0) r^2/(1+z)^2$ remains constant along the route, so
\be \label{F0evo}
    \frac{\partial F^0(z, E_0)}{\partial z} = 2 \frac{F^0(z, E_0)}{1+z} - \frac{2 F^0(z, E_0)}{r\, H} = 2 F(z, E(z)) \left(1 - \frac{1+z}{rH}\right).
\ee
We will also use the following relations, which are direct consequences of Eq.~\eqref{FF0}
\begin{align}
    \frac{\partial F}{\partial z} &= \frac{\partial F^0}{\partial z} \frac{1}{1+z} -\frac{F^0}{(1 + z)^2}\,,\\
    \frac{\partial F}{\partial E} &= \frac{\partial F^0}{\partial E_0}\frac{1}{1+z}\,.
\end{align}
What is relevant for the evolution is the change in $F(z, E(z))$ when $E(z)$ is constant:
\begin{equation}
    \frac{dF}{dz} dz \equiv F(z+dz, E(z)) - F(z, E(z)) = \frac{F^0(z+dz, E_0+dE_0)}{1+z+dz} - \frac{F^0(z, E_0)}{1+z}\,,
\end{equation}
where $E(z) = E_0 (1+z) = (E_0+dE_0)\, (1+z+dz)$ so that
\begin{equation}
    dE_0 = -\frac{E_0 dz}{1+z} = -\frac{Edz}{(1+z)^2}\,.
\end{equation}
We can then write
\begin{equation}
    \frac{dF}{dz} = -\frac{F^0}{(1+z)^2} + \frac{\partial F^0}{\partial z}\, \frac{1}{1+z} + \frac{\partial F^0}{\partial E_0}\, \frac{dE_0/dz}{1+z} = - \frac{F}{1+z} - \frac{\partial F}{\partial E}\, \frac{E}{(1+z)^2} + \frac{\partial F^0}{\partial z}\frac{1}{1+z}\,. \label{NoTildeF}
\end{equation}
Replacing  $\partial F^0/\partial z$ from Eq.~\eqref{F0evo}, we obtain
\be \label{withperr}
    \frac{dF}{dz} = \frac{F}{1+z} - \frac{\partial F}{\partial E}\, \frac{E}{(1+z)^2}-\frac{2F}{r\, H}\,.
\ee
In the presence of inelastic interactions that change the number of particles of a given species, the rates of these processes multiplied by $dt/dz=-1/(H(1+z))$ should be added to the right side of the equation.  

In Eq.~\eqref{withperr}, the third term is proportional to $1/r$. To avoid the pole at $r \to 0$,  let us define
\begin{equation} \label{relation}
    \tilde{F}(z, E) = F(z,E)\, r^2(z)\,,
\end{equation}
for which
\be
    \frac{\partial \tilde{F}^0}{\partial z}=\frac{2 \tilde{F}^0}{1+ z}\,.
\ee
Since the number densities of the neutrino and antineutrino backgrounds, as well as the cross sections of $\nu_i$ and $\bar{\nu}_i$ scattering off the background, are equal (cf. Eqs.~\eqref{sigma}, \eqref{sigmas}, \eqref{nStandard}, and~\eqref{nnn}), the evolution of the fluxes of neutrino and antineutrino as well as their contributions to the ALP flux are similar. Moreover,  IceCube cannot distinguish between neutrinos and antineutrinos. Therefore, it is enough to derive the evolution equation for the sum of neutrino and antineutrino fluxes.
Finally, the evolution formulas for the $\nu\, \nu \to a\, a$ scenario for which the background densities are given by Eqs.~\eqref{Fnui0} and~\eqref{Fa0}.

\section{UV completion of model \label{app:UVcompletion}}
Here, we discuss a possible UV origin of the terms proposed in Eqs.~\eqref{aFF}, \eqref{aanunu}, and~\eqref{aaprimenunu}), one by one. We also show how by coupling to an ultralight dark matter field, certain light particles in our models obtain a varying mass, which prevents the production of extra-relativistic degrees of freedom in the early Universe.

\subsubsection*{\boldmath The $g_{a\gamma \gamma}$ coupling}
The coupling $g_{a\gamma \gamma}$ can come from the axial anomaly as shown in Refs.~\cite{Kim:1979if, Shifman:1979if} (see also Ref.~\cite{DiLuzio:2020wdo} for a review). For example, if a pair of electrically charged chiral fermions ($\Psi_R$, $\Psi_L$) and a scalar singlet $\chi$ transform under a new global $U(1)$ symmetry, like
\begin{equation}
    \Psi_L\to e^{i \alpha/2} \Psi_L \ ,  \ \  \  \Psi_R\to e^{-i \alpha/2} \Psi_R \ \ \    {\rm and} \ \ \ \chi\to e^{i\alpha}\chi\,,
\end{equation} 
then a term $\chi\, \bar{\Psi}_L\, \Psi_R$ is allowed. The ALP can be identified as the Goldstone boson associated with this symmetry
\begin{equation}
    \chi = \frac{v_a + \rho}{\sqrt{2}}\, e^{i\, a/v_a}\,.
\end{equation}
We refrain from calling this $U(1)$ symmetry the Peccei-Quinn symmetry, because we take the particles charged under QCD to be neutral under $U(1)$. Otherwise, an ALP with a relatively large $F_{a\gamma \gamma}$ that is required for our mechanism will obtain too large mass via mixing with pions. The price is that the QCD strong CP problem is not solved in the present setup. 

\subsubsection*{\boldmath The $a^2\nu^Tc\nu$ coupling}
The effective coupling in Eq.~\eqref{aanunu} can originate from an interaction of the following form after integrating out the Majorana neutrino $N_1$:
\be \label{g1}
    g_1 N_1^Tc\nu_\tau a \  .
\ee
Integrating out $N_1$, we obtain
\be
    4 \Lambda_{aa}=\frac{m_{N_1}}{g_1^2} \,.
\ee
Imposing a $U(1)$ flavor symmetry under which the first and second generations are neutral but $\nu_\tau$ (as well as $\tau$) and $N_1$ have opposite charges guarantees the absence of a coupling to $\nu_\mu$ and $\nu_e$.  
As a result, three body decays such as $K^\pm \to \mu^\pm a\, N_1$ or $\pi^\pm \to \mu^\pm a\, N_1$ are forbidden, so we do not need to worry about the bounds from kaon or pion decay.

Within the SM, the conservation of lepton flavor implies that all the $\tau$ decay modes contain $\nu_\tau$ in the final state. In fact, $\tau$ has multiple decay modes ($\tau^-\to X\nu_\tau$ where $X=\mu^- \bar{\nu}_\mu,$ $e^- \bar{\nu}_e,$ $\pi^-,$   $ \pi^- \pi^+\pi^-,$ etc.) with comparable branching ratios which have been measured with a relative accuracy of $\sim 5 \times 10^{-4}$~\cite{ParticleDataGroup:2022pth}. In our model, each decay mode will be accompanied by a new mode replacing $\nu_\tau$ with $N_1\, a$ with $\text{Br}(\tau \to X N_1 a) / \text{Br}(\tau \to X \nu_\tau) \sim g_1^2/(16\pi^2)$. Demanding that this ratio be smaller than the uncertainty in the measurement of $\text{Br}(\tau \to X\nu_\tau)$, we find $g_1<0.3$. To obtain $\Lambda_{aa} \sim 0.5$~GeV, we therefore require $m_{N_1} < 200$~MeV. We take $N_1$ to be heavier than 10~MeV so that for the processes relevant to the present scenario with $s < (2\ {\rm MeV})^2$, using the effective interaction is justified. With $g_1 < 0.3$ and $m_{N_1} > 10$~MeV, we find $\Lambda_{aa}>25$~MeV. As a result, the whole range of our interest (the range $\Lambda_{aa}>465$~MeV which is safe from TXS 0506+056 bound as shown in Eq.~\eqref{optical-depth}) can be easily covered.

The coupling~\eqref{g1} can also lead to lepton number-conserving processes such as $\nu\, \bar{\nu} \leftrightarrow a\, a$ or $a\, \nu\to a\, \nu$ (along with lepton number-violating processes $\nu\, \nu \to a\, a$ or $a\, \nu \to a\, \bar{\nu}$) but their cross sections are further suppressed by $s/m_{N_1}^2$ so we neglect these processes in our analysis. 

The interaction term in Eq.~\eqref{g1} is not invariant under the electroweak symmetry. We suggest the following mechanism to obtain it after the spontaneous electroweak symmetry breaking. Let us suppose we have another sterile neutrino $N_2$ with an electroweak invariant coupling of form\footnote{If we want to also preserve the global $U(1)$ symmetry, we write it as $N_1^T\, c\, N_2\, \chi$, assigning $U(1)$ charge of $-1$ to $N_1$.} 
\be
    g_2\, N_1^T\, c\, N_2\, a\,.
\ee
Then, if $N_2$ mixes with $\nu_\tau$ with a mixing angle $\beta$, we obtain the desired interaction form: $g_1 = g_2\, \sin\beta$. The mixing can be obtained by a Yukawa coupling to the SM Higgs as widely discussed in the literature.

If $N_2$ is heavier than the $Z$ boson, the $Z$ invisible decay width should be suppressed. This mode is measured precisely at LEP~\cite{ParticleDataGroup:2022pth} and recently by CMS~\cite{CMS:2022ett}. 
In our model, $N_2$ decays to $a\, N_1$ and subsequently to $N_1 \to a\, \nu_\tau$, so $N_2$ appears as missing energy at the detectors. If $N_2$ is lighter than $Z$, new invisible $Z$ decay modes $\nu_\tau \bar{N}_2,\bar{\nu}_\tau {N}_2 $ and $\bar{N}_2N_2$ appear along with suppressed $\nu_\tau\bar{\nu}_\tau$. The couplings of $\nu_\tau$ and $N_2$ to $Z$ in our model are equal to the $\nu_\mu$ coupling to $Z$ multiplied respectively by $\cos\beta$ and $\sin \beta$. For $m_{N_2}\ll m_Z$, we can write $\Gamma(Z\to \nu_\tau \bar{\nu}_\tau) = \cos^4 \beta\, \Gamma^\text{SM}(Z\to \nu_\mu \bar{\nu}_\mu) $, $\Gamma(Z\to \nu_\tau \bar{N}_2) = \Gamma(Z\to \bar{\nu}_\tau {N}_2) = \cos^2 \beta \sin^2\beta\, \Gamma^\text{SM}(Z\to \nu_\mu \bar{\nu}_\mu) $ and $\Gamma(Z\to N_2 \bar{N}_2) = \sin^4 \beta\, \Gamma^\text{SM}(Z\to \nu_\mu \bar{\nu}_\mu)$. Thus, up to a correction of $(m_{N_2}/m_Z)^2$, $\Gamma(Z\to {\rm invisibles})$ will be equal to the SM prediction. Therefore, the invisible decay rate of $Z$ cannot constrain our model.
Thus, the invisible decay mode of $Z$ cannot constrain $\beta$. Similar bounds come from various tau decay modes. As long as the mass of $N_2$ is lighter than $m_\tau -m_K$, these limits can be avoided because suppression of standard tau decay modes can be compensated for by decaying into $N_2$. For example, the rate of $\tau \to \nu_\tau\, \pi$ within the SM will be almost equal to the rate of $\tau \to \pi+$invisibles that consists of $\tau \to \pi\, \nu_\tau$ and $\tau \to \pi\, N_2$ which are suppressed, respectively, by $\cos^2\beta$ and $\sin^2 \beta$. Decay modes such as $\tau \to \pi\, a\, N_1$ can also appear along with $\tau \to \pi\, \nu_\tau$, but the rate of this process is suppressed by $(1/16\pi^2)\times g_1^2\sim 10^{-4}$, making them smaller than the measurement uncertainty (for a very recent compilation of the bounds, the reader may consult Ref.~\cite{Blennow:2023mqx}). In principle, the mixing of a sterile neutrino of mass 1 GeV with $\nu_\tau$ is constrained by experiments such as T2K~\cite{T2K:2019jwa} and BEBC WA66~\cite{Barouki:2022bkt}. However, these bounds, which are based on the visible decay of $N_1$, can be relaxed because, in our model, $N_2$ quickly decays into invisible $N_1\, a$, suppressing the visible decay modes of $N_2$ which provide the signal in these experiments.

As discussed in Section~\ref{sec:astro}, as long as $\Lambda_{aa} < 5 \times 10^{5}$~GeV, ALPs can be produced in the early Universe before neutrino decoupling, contributing to the number of relativistic degrees of freedom on which there are strong bounds~\cite{Planck:2018vyg}. Furthermore, as discussed again in Section~\ref{sec:astro} with $\Lambda_{aa}<10^5$~GeV,  neutrinos cannot freely stream at $z \sim 10^5$ which is in tension with CMB observations~\cite{Taule:2022jrz}.
In the literature, it has been shown that such bounds can be relaxed by invoking non-standard cosmologies~\cite{Allahverdi:2020bys} or varying mass of $N_1$ through coupling to the environment~\cite{Hannestad:2013ana, Dasgupta:2013zpn, Chu:2015ipa, Chu:2018gxk, Paul:2018njm, Farzan:2019yvo}. As an example in the following, we propose a mechanism to avoid these bounds by varying the effective mass of $N_1$ by coupling it with ultralight dark matter, $\phi_\text{dm}$:
\be \label{DMNN}
    \frac{|\phi_\text{dm}|^2}{\mathcal{M}}\, N_1^T\, c\, N_1\,,
\ee
where $\mathcal{M}$ is a constant with a dimension of mass and 
\be \label{rhoCLASSIC}
    |\phi_\text{dm}|^2=\frac{\rho_\text{dm}}{m_\text{dm}^2}\ .
\ee
Thus, the varying mass of $N_1$ will be $m_{N_1}+\rho_\text{dm} / (m_\text{dm}^2 \mathcal{M})$. At $z=10^5$, $\rho_\text{dm}=z^3 \rho_\text{dm}^0 =10^4$~eV$^4$. With $m_\text{dm}^2 \mathcal{M}<10^{-11}$~eV$^3$, we find that $N_1$ becomes heavier than $5\times 10^5$~GeV at $z=10^5$ which implies that neutrinos can freely stream at the recombination and, moreover, ALPs cannot be produced in the early Universe.

Through the coupling with $\nu_\tau$, the lepton $N_1$ can be produced and thermalized inside the supernova core. Moreover, as shown in Ref.~\cite{Cerdeno:2023kqo, Fiorillo:2023cas, Fiorillo:2023ytr} contrary to some claims, within this regime with large coupling, the secret interaction of neutrinos cannot change the duration of neutrino bursts. However, $N_1N_1 \to \phi_\text{dm} \phi_\text{dm}$ may lead to fast cooling of supernovae. To avoid it, we should ensure that the rate of $\phi_\text{dm}$ production inside the supernova core per each $N_1$ (that is, $\sigma (N_1N_1 \to \phi_\text{dm} \phi_\text{dm})\, n_{N_1}$) is much lower than the rate of energy loss of supernovae, $10^{-1}$ sec$^{-1}$. Taking the thermal density for $N_1$ with $T\sim 30$~MeV and $\sigma(N_1 N_1 \to \phi_\text{dm} \phi_\text{dm})\sim 1/(4\pi \mathcal{M}^2)$, we find that $\mathcal{M}$ should be much larger than $10^9$~GeV to avoid the bounds from supernova energy loss.
Taking, e.g. $\mathcal{M}=10^{10}$ GeV,  with $m_\text{dm}\sim   10^{-15}$~eV we can have $\mathcal{M}\, m_\text{dm}^2=10^{-11}~{\rm eV}^3$.
With these parameters, the contribution from Eq.~\eqref{DMNN}  to the $N_1$ mass at the low redshift in the intergalactic space is negligible.
That is, taking $\rho_\text{dm}\sim 1~{\rm GeV}/{\rm cm}^3$, the correction to $N_1$ is less than 1~MeV and much smaller than the bare mass of $N_1$.
At temperatures above MeV, the effective $N_1$ mass will far exceed the PeV, so the ALP and $N_1$ cannot be produced before neutrino decoupling, maintaining the SM prediction for the effective relativistic degrees of freedom. At temperatures above 1~GeV, $N_2$ is produced by mixing with $\nu_\tau$ but decays to SM light fermions long before the neutrino decoupling era.

\subsubsection*{\boldmath The $a a'\nu^Tc\nu$ coupling}
If in addition to the Yukawa coupling shown in Eq.~\eqref{g1} there is an interaction of the form $N_1^T\, c\, \nu_\tau\, a'$, we obtain the $\Lambda_{aa'}$ coupling in Eq.~\eqref{aaprimenunu} along with $\Lambda_{aa}$. However, we need a mechanism that forbids the latter. This can be achieved by introducing an electroweak singlet scalar with couplings of the form 
\be
    \lambda_1\, \nu_\tau^T\, c\, \nu_\tau\, \phi + \lambda_2\, m_\phi\, \phi\, a\, a'\,.
\ee
Integrating out $\phi$, we obtain the $\Lambda_{aa'}$ coupling without inducing the $\Lambda_{aa}$ coupling. We may assign tau lepton numbers of 2 and $-2$, respectively, to the fields $a'$ and $\phi$ to forbid the coupling $\phi a^2$. Moreover, the same symmetry forbids a coupling between $\phi$ and $\nu_e$ or $\nu_\mu$. While the term $\lambda_2$ is electroweak invariant, $\lambda_1$ can appear only after the electroweak symmetry breaking. This coupling can be obtained by mixing $\phi$ with a scalar electroweak triplet of the tau lepton number of two that couples to a pair of $(\nu_\tau\, \tau)$ doublets. There are no bounds on $\lambda_2$, so it can be as large as 1. A bound of 0.02 on $\lambda_1$ is derived by Ref.~\cite{Chang:2022aas} from supernovae, but Refs.~\cite{Fiorillo:2023cas, Fiorillo:2023ytr} refute this bound. With $\lambda_1\lambda_2\sim 0.01$, to obtain $\Lambda_{aa'} \sim 1000$~MeV, $m_\phi$ should be of order of 10~MeV. Through the $\lambda_1$ coupling, $\phi$ can be produced in the early Universe but as long as $m_\phi > 10$~MeV, it can decay to $\nu_\tau\, \nu_\tau$ before the BBN era. We should, however, introduce a mechanism to prevent the $a'$ and $a$ production in the early Universe. Again, this can be achieved by coupling to ultralight dark matter:
\begin{equation}
    \lambda_3\, |\phi_\text{dm}|^2\, |a'|^2,
\end{equation}
which induces an effective mass of $\sqrt{\lambda_3} \sqrt{\rho_\text{dm}}/m_\text{dm}$ for $a'$.

As discussed about $N_1$ above, $\phi$ cannot change the supernova observables because it quickly decays to particles that thermalize with neutrinos. However, the $\lambda_3$ coupling can lead to $\phi_\text{dm}$ production via $a'a'\to \phi_\text{dm}\phi_\text{dm} $ and therefore to fast supernova energy loss, unless $\sigma(a'a'\to \phi_\text{dm}\phi_\text{dm})\, n_{a'} \ll (10~{\rm sec})^{-1}$ where 10~sec is the typical duration of the $a'$ stay in the thermal bath of the protoneutron star. Taking $n_{a'}=1.2\, T^3/\pi^2$, $\sigma\sim \lambda_3^2/(4\pi T^2)$ and $T\sim 30$~MeV, this implies $\lambda_{13}\ll 10^{-11}$.

The cross section of ALP production via $\phi\, \bar{\phi} \to a\, a$ in the early Universe with $T > m_\phi$ is given by $\sim \lambda_2^4\, m_\phi^4/(4\pi\, m_{a'}^4\, T^2)$. To prevent the thermal production of ALPs at $T > \mathcal{1}$~MeV, $m_{a'}$ should be larger than 2.5~TeV. On the other hand, the effective mass of $a'$ in the space between galaxies today with $\rho_\text{dm} = 1.3\times 10^{-6}$~GeV$/{\rm cm}^3$ should be smaller than keV, which implies $m_\text{dm}/\sqrt{\lambda_3} > 3 \times 10^{-9}$~eV. With $m_\text{dm}/\sqrt{\lambda_3} = 10^{-8}$~eV, both conditions will be satisfied. 
Taking $\lambda_3\sim 10^{-12}$ which respects the supernova energy-loss bounds, $m_\text{dm}/\sqrt{\lambda_3}\sim 10^{-8}$~eV implies $m_\text{dm}\sim 10^{-14}$~eV.

Furthermore, for temperatures higher than the $a'$ vacuum mass, the effective mass of $a'$ will be too large to be thermally produced. Thus, the contribution of ALP and $a'$ and of $\phi$ to the effective degrees of freedom will be negligible and the background relic ALP will be irrelevant to our solution.

\section{ALP-photon conversion\label{app:ALPconversion}}
Let us now discuss the photon flux at lower energy ranges that are measured by LHAASO and Fermi-LAT. The Hamiltonian governing the ALP photon conversion in the $(A_x,\, A_y,\, a)$ basis can be written as 
\begin{equation}
    \mathcal{H}=\left(
    \begin{matrix}
        \Delta_x &0 & \Delta_{a\gamma_{x}}\\
        0& \Delta_y  & \Delta_{a\gamma_{y}}\\
        \Delta_{a\gamma_{x}}& \Delta_{a\gamma_{y}}& \Delta_a
    \end{matrix}
    \right).
\end{equation} 
The off-diagonal elements of the Hamiltonian are independent of the ALP energy
\begin{equation}
    \Delta_{a \gamma_i } =\frac{g_{a\gamma\gamma}}{2}\, B_i = 5 \times 10^{-3} \left(\frac{g_{a\gamma\gamma}}{3\times 10^{-12}~{\rm GeV}^{-1}}\right) \frac{B_i}{10^{-6}~{\rm G}} {\rm kpc}^{-1}.
\end{equation}
For an ultrarelativistic ALP, $\Delta_a = m_a^2/(2E_a)$. However, for low energies, $\Delta_a$ is larger than the rest of the elements. For the range of the sensitivity of Fermi-LAT-- which is $[0.1,\, 1]$~GeV-- the ALP-photon mixing is suppressed to $\Delta_{a \gamma_{x,y}}/\Delta_a < 6 \times 10^{-6}$ so the ALP to photon conversion probability cannot exceed $10^{-10}$. With this probability, the contribution of our model, the photon flux at $0.1-1$~GeV will be more than 10 orders of magnitude below the Fermi-LAT photon flux measurement. 

LHAASO has also searched for photons in the energy range of 0.5~TeV to 10~TeV, finding more than 5000 events. We should make sure that our prediction for photons does not exceed this observation. The number of ALPs with energy in this range is larger than that with energy higher than 10~TeV by a factor of just 20. Recalling that the LHAASO has detected only one event with energy above 10~TeV, with  $P_{a\to \gamma}$ and an effective area constant against energy, we would therefore expect about 20 events from our model within $[0.5,\, 10]$~TeV which can easily hide in the statistical fluctuation of the number of observed events. Considering that both $P_{a\to \gamma}$ and the effective area decrease with the energy, the contribution from our model will be completely negligible, so our model is unconstrained by the LHAASO low-energy range data.

\bibliographystyle{JHEP}
\bibliography{biblio}
\end{document}